\documentclass[12pt]{iopart}
\usepackage{iopams}
\usepackage{ifpdf}
\usepackage{amssymb}
\usepackage{txfonts}
\usepackage{graphicx}
\usepackage{subfigure}
\usepackage{dcolumn}
\usepackage{bm}
\usepackage{sidecap}
\usepackage{enumitem}
\usepackage{cite}
\usepackage{color}

\begin{document}

\title{Realization of phase locking in good-bad-cavity active optical clock}

\author{Tiantian Shi,$^{1}$ Duo Pan,$^{1,2}$ and Jingbiao Chen$^{1}$}

\address{$1$ State Key Laboratory of Advanced Optical Communication, System and Network,School of Electronics Engineering and Computer Science, Peking University,Beijing 100871, China}
\address{$2$ Author to whom any correspondence should be addressed.}

\ead{duopan@pku.edu.cn}
%\thanks{these authors contributed equally to this work}
\vspace{10pt}
\begin{indented}
\item[]May 2019
\end{indented}

\begin{abstract}
The residual cavity-pulling effect limits further narrowing of linewidth in dual-wavelength (DW) good-bad-cavity active optical clocks (AOCs). In this paper, we for the first time experimentally realize the cavity-length stabilization of the 1064/1470 nm DW-AOCs by utilizing the phase locking technique of two independent 1064 nm good-cavity lasers. The frequency tracking accuracy between the two main-cavities of DW-AOCs is better than $3 \times {10^{ - 16}}$ at 1 s, and can reach $1 \times {10^{ - 17}}$ at 1000 s. Each 1470 nm bad-cavity laser achieves a most probable linewidth of 53 Hz, which is about a quarter of that without phase locking. The influence of the asynchronous cavity-lengths variation between two DW laser systems is suppressed.
\end{abstract}

%%%%%%%%%%%%%%%%%%%%%%%%%%  body  %%%%%%%%%%%%%%%%%%%%%%%%%%
\section{Introduction}
Optical clocks~\cite{Nature_506_12941_(2014),Nature_Communications_vol_6_6896_(2015),Nature_Photonics9_185_189(2015),Science_358_90_94_(2017),Nature_564_87_90(2018),Phys.Rev.Lett.104.070802.2010,	Phys.Rev.Lett.116.063001.2016,Phys.Rev.A.99.011401(R).2019}, whose frequencies are about five orders of magnitude larger than the microwave atomic clocks, have potential better stability and accuracy than that of the cesium (Cs) fountain microwave clocks used to define the second~\cite{Metrologia_42_64_2005, Metrologia_48_283_289_2011, Metrologia_53_1123_1130_2016, Metrologia_55_789_805_2018}. Thus they attract a lot of interest in precision measurements~\cite{Reviews_of_Modern_Physics_2015_87(2)_637}.

Most optical clocks are based on the passive mode of operation, and the extremely coherent sources is necessary in the advanced optical clocks for interrogating ultranarrow atomic transitions with linewidths of a few mHz. Therefore, the heart of the passive optical clocks is the local oscillator stabilized to a high-finesse optical reference cavity~\cite{NaturePhotonics6_687_692(2012)}. Tremendous progress has been achieved in the filed of passive optical clocks recent years, and current state of the art neutral atom optical lattice clocks~\cite{Nature_506_12941_(2014),Nature_Communications_vol_6_6896_(2015),Nature_Photonics9_185_189(2015),Science_358_90_94_(2017),Nature_564_87_90(2018)} and single-ion optical clocks~\cite{Phys.Rev.Lett.104.070802.2010,	Phys.Rev.Lett.116.063001.2016,Phys.Rev.A.99.011401(R).2019} have achieved the frequency stabilities of ${\rm{1}}{{\rm{0}}^{{\rm{ - 18}}}}$ even ${\rm{1}}{{\rm{0}}^{{\rm{ - 19}}}}$ by adopting the ultra-stable optical cavities running in the room temperature. However, the improvements of the instability of passive optical clocks now are limited by the Brownian thermomechanical noise of the reference cavity. To further reduce the thermal noise of the reference cavity and improve the instability of the interrogation lasers, the reference cavity is usually cooled down to a low temperature~\cite{NaturePhotonics6_687_692(2012),Phys_Rev_Lett_vol_118_no_26_p_263202_(2017),Optica.6.240.243.2019}, which will inevitably increase the complexity of realizing the frequency standards with ultra-narrow linewidth.

As the counterpart, the active optical clocks (AOCs), with the superiorities of weaker cavity-pulling effect and narrower quantum-noise-limited linewidth in bad-cavity regime, were proposed~\cite{2005IEEE_Int_Frequency_Control_Symp,Chin_Sci_Bull_54(3)2009,Phys.Rev.Lett.102.163601.2009}. Unlike the traditional passive optical clocks, the AOC works in the bad-cavity regime with the linewidth of cavity mode $\Gamma_{\rm{cavity}}$ being much wider than that of the laser gain $\Gamma_{\rm{gain}}$. Therefore, the laser frequency of the AOC will not follow the fluctuations of the cavity length exactly, but in a form of the suppressed cavity-pulling effect, and the extent of the suppression equals the bad-cavity coefficient $a$, where $a = {{{\Gamma_{\rm{cavity}}}} \mathord{\left/{\vphantom {{{\Gamma _{\rm{cavity}}}} {{\Gamma _{\rm{gain}}}}}} \right.\kern-\nulldelimiterspace} {{\Gamma _{\rm{gain}}}}}$. Since it was proposed, various types of active atomic systems have been investigated~\cite{Nature.48478.2012,Chin_Sci_Bull_58_pp_2033_2038_2013,Opt_Lett_39_6339_(2014),Optics_Express22_pp13269_13279(2014),IEEE_IFCS_EFTF_2015_PP363_368,Phys_Rev_A_96_023412(2017),Physical_Review_A96_013847(2017),PRX8_021036(2018)}. Recently, a superradiant pulse from the strontium clock transition with a fractional Allan deviation of $6.7(1) \times {10^{ - 16}}$ at 1 s has been realized \cite{PRX8_021036(2018)}. In our lab, a continuous-wave cesium (Cs) 1470 nm bad-cavity laser with the linewidth of 149 Hz  \cite{IEEE_TUFFC_65PP1958_1964(2018)}, which is much narrower than the 1.81 MHz natural linewidth of 1470 nm transition \cite{IEEE_IFCS_2014pp242_245}, has been achieved. However, without any cavity-length stabilization, the linewidth of 1470 nm laser is still influenced by the residual cavity-pulling effect, which limits the applications of the AOCs as a high-precision frequency standard.

For further reducing the residual cavity-pulling effect in Cs four-level AOC and to achieve a continuous active optical frequency standard with high performances, we have proposed the 1064/1470 nm dual-wavelength (DW) good-bad cavity AOC, which is described in \cite{IEEE_TUFFC_65PP1958_1964(2018),Chin_Phys_Lett_32_083201(2015),Rev_Sci_Inst_89_043102(2018)}.  The idea of DW-AOC is to make the good-cavity laser and the bad-cavity laser share a single main-cavity, and the residual cavity-pulling effect of the bad-cavity laser can be effectively suppressed by stabilizing the cavity length with the good-cavity laser, due to the fact that the bad-cavity laser is insensitive to the cavity-puling effect. The core part, the DW good-bad-cavity system, has recently been realized in~\cite{IEEE_TUFFC_65PP1958_1964(2018)}, and the performance of the 1064/1470 nm DW-lasers was also analyzed. However, even previous works experimentally proved that the 1470 nm bad-cavity laser has the advantage of the suppression of the cavity-pulling effect, the analysis shown that the residual cavity-pulling effect was still not negligible, which will affect the output linewidth of the 1470 nm laser. If the residual cavity-pulling effect cannot be effectively suppressed, it is hard to analyze and then suppress the other technical noises of DW-AOC system in experiments, which affects the optimization of the system and limit its applications.

In this paper, we focus on the experimental realization of the phase locking and the performance optimization in the DW good-bad-cavity laser systems, to suppress the residual cavity-pulling effect, which is a necessary step moving towards the cavity-stabilized AOCs. We built two independent DW systems and locked the two cavities together by phase locking technique~\cite{Appl_Phys_B_vol_60_S241_S248_1995,Science_288_635_699_(2000),Optics_Letters_34_2958_2960_(2009),OpticsExpress_vol18_pp8621-8629_2010,IEEE_Photonics_Technology_Letters_Vol_24_2012,Appl_Phys_B_vol_123_no_9_2017} of 1064 nm good-cavity lasers, which can synchronize the cavity-lengths change between two DW-AOCs. It is aimed at eliminating the impact of the common-mode noise, which is caused by the asynchronous lengths variation of two independent main-cavities of DW-AOCs, on the beating linewidth of the 1470 nm bad-cavity lasers. Four key techniques are exploited in this work. First, we improve the performance of the pumping laser by the narrow-linewidth interference filter configuration external cavity diode laser (IF-ECDL), whose linewhidth is smaller than 20 kHz. Second, the modulation transition spectroscopy (MTS) technique is exploited to further optimaize the frequency stability of the 459 nm pumping laser. Then, the optical phase-locked loop (OPLL) technique is used to achieve the phase locking of 1064 nm lasers, together with a divider to increase the robustness of the 1064nm beating signal to the influence of external noises. Finally, the Cs microwave atomic clock is used as the reference source to improve the tracking accuracy between two main-cavities of DW systems.

The experiment results show that, the beating linewidth of 1064 nm good-cavity lasers is narrower than 1 Hz after phase locking, which is limited by the resolution bandwidth (RBW) of the spectrum analyzer. Furthermore, the fractional frequency stability of the offset frequency of 1064 nm lasers is measured by a frequency counter and calculated by the Allan deviation, which indicates that the tracking accuracy between two DW systems is better than ${\rm{3}} \times {\rm{1}}{{\rm{0}}^{{\rm{ - 16}}}}$ at 1 s, and can reach $1 \times {10^{ - 17}}$ at 1000 s. The most probable linewidth of each 1470 nm bad-cavity laser is narrowed from 223 Hz to 53 Hz after cavity-length stabilization. The causal analysis and optimization methods are also performed.      

This paper is organized as follows: In Section~\ref{exprimental_part}, the experimental setup and the phase locking process are introduced. Section~\ref{Results_and_discussions} gives the experimental results and the analysis of the experiment, and the conclusion is drawn in Section~\ref{conclusion}.

\section{Experiment}
\label{exprimental_part}

\subsection{Wavelengths choice of DW lasers}
Before we begin to our practical experiment, we first give the explanation of why the 1470 nm and the 1064 nm are chosen to be the wavelengths of the bad- and good- cavity lasers, respectively.

In general, the atomic energy level of the Cs four-level AOC is shown in Fig.~\ref{energy-level}. The pumping and lasing process of the four-level system can be described as follows:

\begin{figure}[h]
	\centering
	\subfigure[]{\includegraphics[width=0.45\textwidth]{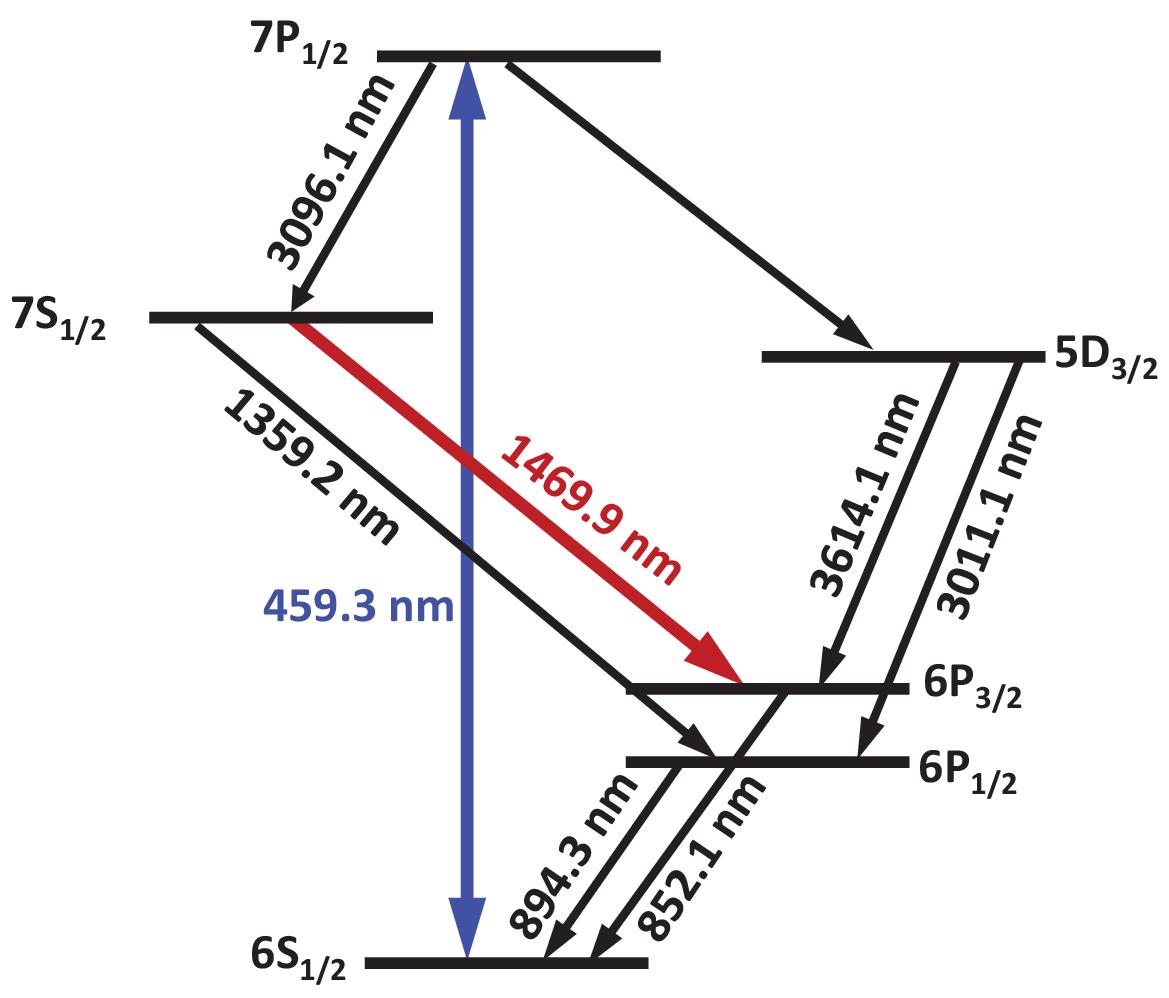}
		\label{energy-level} }
	\subfigure[]{\includegraphics[width=0.45\textwidth]{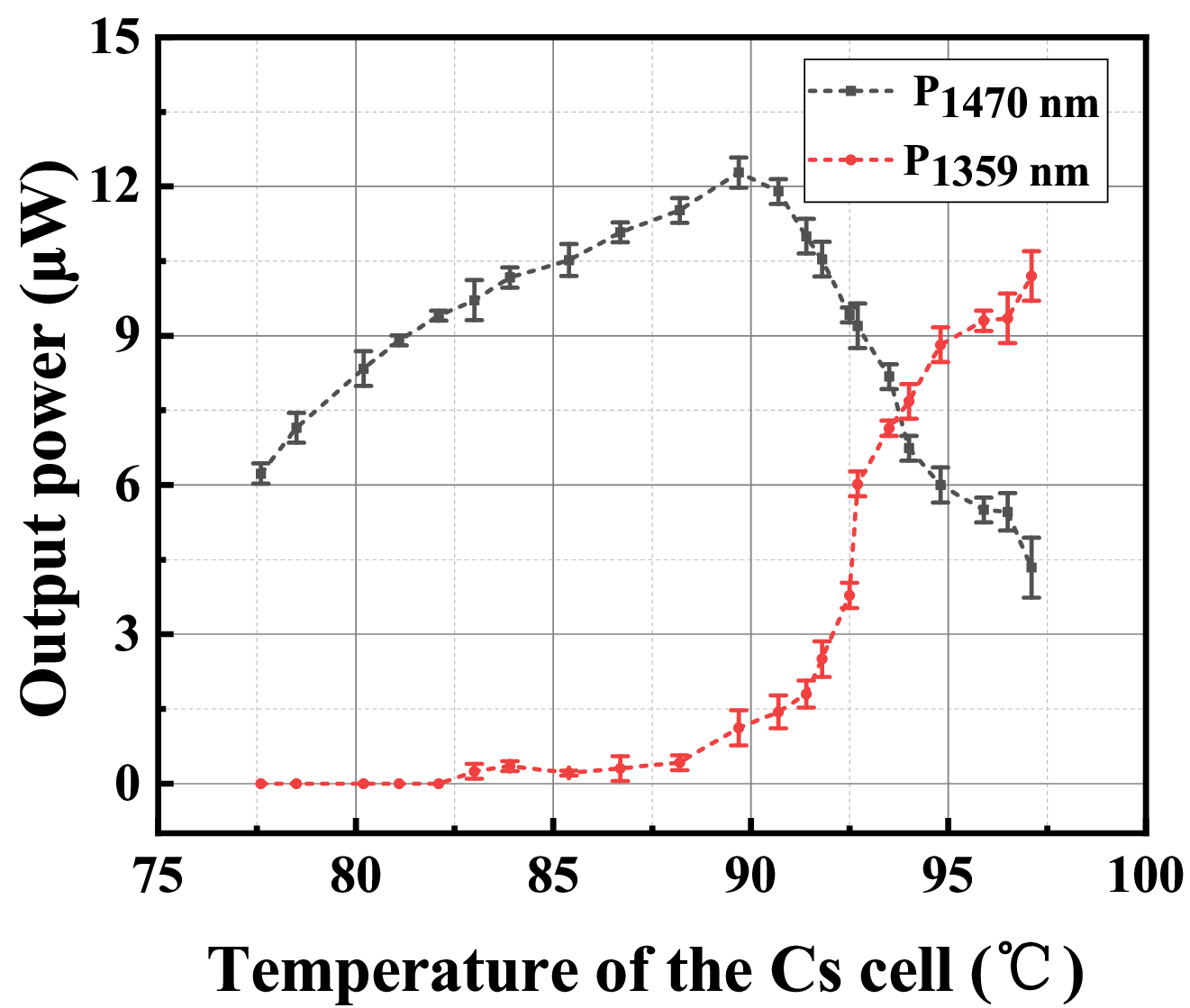}
		\label{P1470-P1359-Tem-2} }
	\caption{(a) The energy level of Cs four-level active optical clock system. (b) The relation between the temperature of the Cs atomic vapor cell and the power of output lasers in the Cs four-level AOC. The black square-dots and the red square-dots represent the output power of 1470 nm laser and 1359 nm laser, respectively.}
\end{figure}

First, the Cs atoms are pumped from the ${\rm{6}}{S_{{1 \mathord{\left/{\vphantom {1 2}} \right.\kern-\nulldelimiterspace} 2}}}$ ground state to the ${\rm{7}}{P_{{1 \mathord{\left/{\vphantom {1 2}} \right.\kern-\nulldelimiterspace} 2}}}$ state by the 459 nm laser, and then decay to the $7{S_{{1 \mathord{\left/{\vphantom {1 2}} \right.\kern-\nulldelimiterspace} 2}}}$ and $5{D_{{3 \mathord{\left/{\vphantom {3 2}} \right.\kern-\nulldelimiterspace} 2}}}$ states by spontaneous transitions. Second, the atoms at ${\rm{7}}{S_{{1 \mathord{\left/{\vphantom {1 2}} \right.\kern-\nulldelimiterspace} 2}}}$ and $5{D_{{3 \mathord{\left/{\vphantom {3 2}} \right.\kern-\nulldelimiterspace} 2}}}$ states decay to ${\rm{6}}S{}_{{1 \mathord{\left/{\vphantom {1 2}} \right.\kern-\nulldelimiterspace} 2}}$, ${\rm{6}}{P_{{1 \mathord{\left/{\vphantom {1 2}} \right.\kern-\nulldelimiterspace} 2}}}$ and ${\rm{6}}{P_{{3 \mathord{\left/{\vphantom {2 2}} \right.\kern-\nulldelimiterspace} 2}}}$ states, respectively. Third, the atoms at ${\rm{6}}{P_{{1 \mathord{\left/{\vphantom {1 2}} \right.\kern-\nulldelimiterspace} 2}}}$ and ${\rm{6}}{P_{{3 \mathord{\left/{\vphantom {2 2}} \right.\kern-\nulldelimiterspace} 2}}}$ states return to ground state ${\rm{6}}{S_{{1 \mathord{\left/{\vphantom {1 2}} \right.\kern-\nulldelimiterspace} 2}}}$ and then pumped to ${\rm{7}}{P_{{1 \mathord{\left/{\vphantom {1 2}} \right.\kern-\nulldelimiterspace} 2}}}$ state again by pumping laser. In the steady state, the populations of Cs atoms at ${\rm{6}}{S_{{1 \mathord{\left/{\vphantom {1 2}} \right.\kern-\nulldelimiterspace} 2}}}$, ${\rm{7}}{P_{{1 \mathord{\left/{\vphantom {1 2}} \right.\kern-\nulldelimiterspace} 2}}}$, ${\rm{7}}{S_{{1 \mathord{\left/{\vphantom {1 2}} \right.\kern-\nulldelimiterspace} 2}}}$, $5{D_{{3 \mathord{\left/{\vphantom {3 2}} \right.\kern-\nulldelimiterspace} 2}}}$, ${\rm{6}}{P_{{3 \mathord{\left/{\vphantom {3 2}} \right.\kern-\nulldelimiterspace} 2}}}$ and ${\rm{6}}{P_{{1 \mathord{\left/
				{\vphantom {1 2}} \right.\kern-\nulldelimiterspace} 2}}}$ states are $26.1\%$, $25.6 \%$, $5.1 \%$, $38.7 \%$, $1.9 \%$ and $2.4 \%$, respectively. Therefore, the population inversion can be established between $7{S_{{1 \mathord{\left/{\vphantom {1 2}} \right.\kern-\nulldelimiterspace} 2}}} - 6{P_{{3 \mathord{\left/{\vphantom {3 2}} \right.\kern-\nulldelimiterspace} 2}}}$, $5{D_{{3 \mathord{\left/	{\vphantom {3 2}} \right.\kern-\nulldelimiterspace} 2}}} - 6{P_{{3 \mathord{\left/	{\vphantom {3 2}} \right.\kern-\nulldelimiterspace} 2}}}$, $5{D_{{3 \mathord{\left/{\vphantom {3 2}} \right.\kern-\nulldelimiterspace} 2}}} - 6{P_{{1 \mathord{\left/{\vphantom {1 2}} \right.\kern-\nulldelimiterspace} 2}}}$ and $7{S_{{1 \mathord{\left/{\vphantom {1 2}} \right.\kern-\nulldelimiterspace} 2}}} - 6{P_{{1 \mathord{\left/{\vphantom {1 2}} \right.\kern-\nulldelimiterspace} 2}}}$ states.

In this work, we choose the $7{S_{{1 \mathord{\left/{\vphantom {1 2}} \right.\kern-\nulldelimiterspace} 2}}} - 6{P_{{3 \mathord{\left/{\vphantom {3 2}} \right.\kern-\nulldelimiterspace} 2}}}$ transition as the clock transition instead of the others, which corresponds to the 1470 nm laser. The reasons are given below:

For the $5{D_{{3 \mathord{\left/	{\vphantom {3 2}} \right.\kern-\nulldelimiterspace} 2}}} - 6{P_{{3 \mathord{\left/	{\vphantom {3 2}} \right.\kern-\nulldelimiterspace} 2}}}$ and $5{D_{{3 \mathord{\left/{\vphantom {3 2}} \right.\kern-\nulldelimiterspace} 2}}} - 6{P_{{1 \mathord{\left/{\vphantom {1 2}} \right.\kern-\nulldelimiterspace} 2}}}$ transitions, the wavelengths of the these transitions are 3614 nm and 3011 nm, respectively. The frequencies of these two transition lasers are both smaller than that of the 1470 nm transition, which limits their superiorities in optical frequency standards. For the $7{S_{{1 \mathord{\left/{\vphantom {1 2}} \right.\kern-\nulldelimiterspace} 2}}} - 6{P_{{1 \mathord{\left/{\vphantom {1 2}} \right.\kern-\nulldelimiterspace} 2}}}$ transition, the wavelength of this transitions is 1359 nm. It seems that this transition can also be used as the clock transition, so we have also experimentally realized it and the relation between the temperature of the Cs atomic vapor cell and the power of output lasers is also given in Fig.~\ref{P1470-P1359-Tem-2}. It can be seen that, compared to the 1470 nm laser, the 1359 nm laser requires higher temperature of the atomic gain medium to obtain its optimal output power, which will inevitably introduce greater collision shift. Therefore, this transition is not the best choice in our system either. In summary, we finally chose the 1470 nm as the clock transition laser, namely, the bad-cavity laser.

As for the wavelength choice of good-cavity laser, we calculate that one of the magic wavelengths of the $7{S_{{1 \mathord{\left/{\vphantom {1 2}} \right.\kern-\nulldelimiterspace} 2}}} - 6{P_{{3 \mathord{\left/{\vphantom {3 2}} \right.\kern-\nulldelimiterspace} 2}}}$ 1470 nm transition is 1030 nm~\cite{Phys.Rev.A.94.012505.2016}, which is very close to the wavelength of the good-cavity laser, namely, 1064 nm. Thus we choose the Nd: YAG structure to generate the good-cavity signal, avoiding the influence of the light shift caused by the good-cavity laser to the 1470 nm clock transition laser. The beam intensity of the 1064 nm good-cavity laser inside the cavity is 17.8 mW/mm$^2$, and we calculated the light shift of the 1470 nm laser induced by the 1064 nm laser, being 9.62 Hz. Also, the Allan deviation of the 1064 nm laser power is ${\rm{3}} \times {\rm{1}}{{\rm{0}}^{{\rm{ - 5}}}}$ at 1 s in our experiment. Therefore, the linewidth broadening of the 1470 nm bad-cavity laser caused by the light shift of the 1064 nm laser is smaller than 0.3 mHz, which is negligible. Moreover, we also experimentally observed that the center frequency of the 1470 nm beat-note signal hardly varies with the power fluctuation of the 1064 nm laser, which verifies our theoretical analysis. Therefore, we chose the 1064 nm as the good-cavity laser.

\subsection{Experimental setup and the phase locking process}

\begin{figure}[htbp]
	\centering
	\includegraphics[width=11cm]{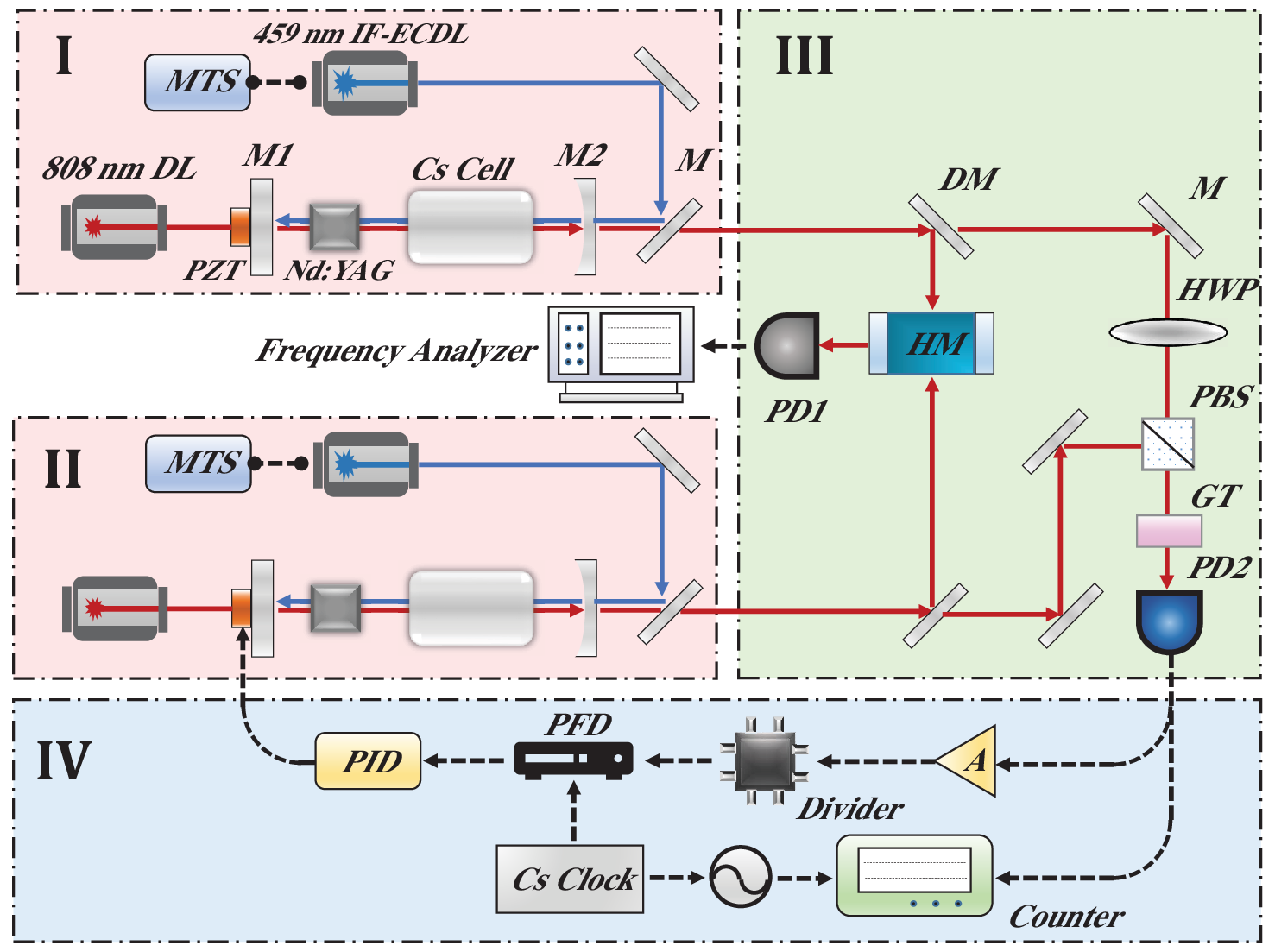}\\
	\caption{Experimental setup of the heterodyne optical phase-locked loop between two DW-AOCs. 808 nm DL, 808 nm diode laser;  459 nm IF-ECDL, 459 nm interference filter configuration external cavity diode laser; MTS, modulation transition spectroscopy; HWP, halfwave plate; PBS, polarized beam splitter; DM, dichroic mirror; HM, 1470 nm heterodyne module; PD, photodetector; GT, Glan-Taylor; A, amplifier; PFD, phase-frequency detector; PID, proportion-integral-derivative locking system; Cs clock, Cs microwave atomic clock.}
	\label{experimentsetup}
\end{figure}

The block diagram in Fig. \ref{experimentsetup} shows the experimental setup of the heterodyne OPLL between two DW-AOCs. For simplicity, we divide our experimental setup into four parts, in which the red boxes (I, II) denote two identical DW laser systems, the green box (III) denoting the beat detection unit of the systems, and the blue box (IV) denoting the phase locking part of our experiment.

\begin{figure}[h]
	\centering
	\includegraphics[width=8cm]{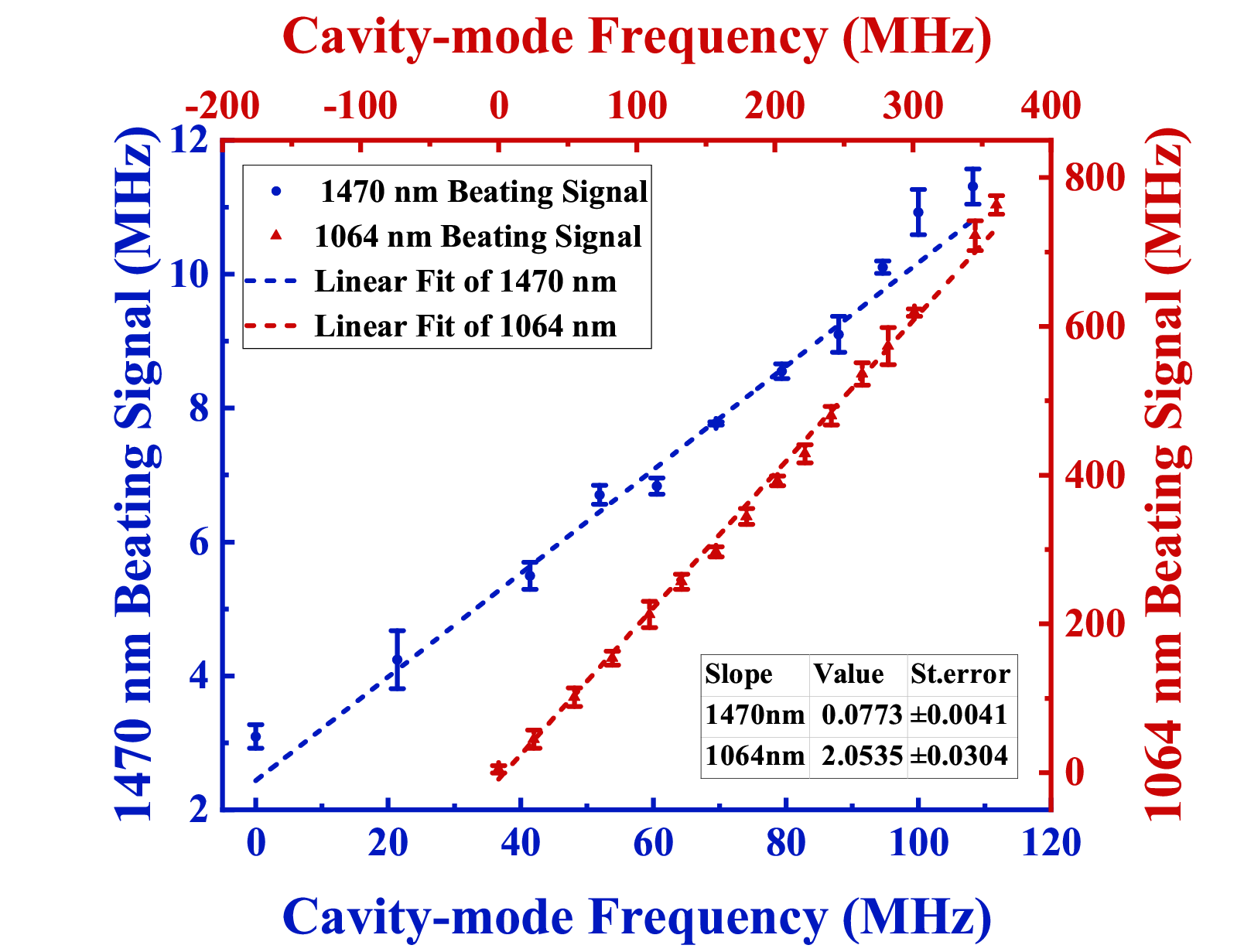}\\
	\caption{The frequency detuning of the 1064/1470 nm DW signals versus that of the cavity mode. The red triangular-dots and the blue circular-dots represent the frequency detuning of the 1064 nm good-cavity laser and the 1470 nm bad-cavity laser. The fitted slop of the frequency detuning of the 1064 nm laser k$_{1}$ is 2.0535 $\pm$ 0.0304, and the fitted slop of the 1470 nm laser k$_{2}$ is 0.0773 $\pm$ 0.0041, which means that the measured bad-cavity coefficient of the DW-AOC is in the interval $a \in \left( {26.548 - 1.713,26.548 + 1.907} \right)$.}\label{1064-1470-cavity-pulling-error-bar}
\end{figure}

In the I and II parts, we build two identical DW systems for analyzing the characteristics of the DW signals by optical heterodyne. The Nd:YAG 1064 nm laser and the Cs 1470 nm laser share a single main-cavity of the DW-AOC and work in the good- and bad-cavity regime by designing cavity-mirrors coating for specific reflectivity at four wavelengths (808 nm, 1064 nm, 459 nm, 1470 nm). The $\Gamma_{\rm{cavity}}$ and the $\Gamma_{\rm{gain}}$ of the 1064 nm good-cavity laser are 100 MHz and 132 GHz respectively, while that of the 1470 nm bad-cavity laser are 288 MHz and 10 MHz, which means that the impact of the cavity-length noise on the 1470 nm laser can be, in principle, reduced by 1/29 compared with the unity for the 1064 nm laser. A 459 nm IF-ECDL, which is stabilized by the MTS technique, pumps the Cs atoms to generate the 1470 nm clock transition laser. The Cs atomic vapor cells are heated to 90 $^{o}$C for increasing the atomic density. Meanwhile, the 808 nm continuous-wave diode laser focus on the Nd:YAG crystal to output the 1064 nm good-cavity laser. Consequently, a single-mode 1064 nm and 1470 nm DW lasers generated from the same cavity simultaneously, and separately work in good- and bad-cavity regime. By observing the center-frequency movement of beat-note signals of 1470 nm lasers and 1064 nm lasers when tuning the piezoelectric ceramics (PZT) of one of the main-cavities, the bad-cavity coefficient of the DW-AOC is measured with uncertainties, as shown in Fig.~\ref{1064-1470-cavity-pulling-error-bar}. We find that the measured data can fit the linear model well. Considering the uncertainty of the measurement results, we obtain the measured bad-cavity coefficient of the DW-AOC is in the interval $a \in \left( {26.548 - 1.713,26.548 + 1.907} \right)$. It means that the frequency detuning of the 1064 nm good-cavity laser changed exactly with the cavity mode, while for the 1470 nm bad-cavity laser, the cavity-pulling effect is suppressed to a large extent.

In the III part, the 1064/1470 nm DW lasers output from each main-cavity are separated through a DM, and then the two 1064 nm beams and two 1470 nm beams are respectively overlapped for optical heterodyning. The beat-note signal of 1470 nm bad-cavity lasers is led to a spectrum analyzer (Agilent N9020A) to measure its linewidth. The beat-note signal of 1064 nm good-cavity lasers is divided into two parts through a directional coupler (TC-2080-10S). One part is sent to the frequency counter (Agilent 53220 A) to record the variation of frequency difference of 1064 nm lasers. The other part is used in the heterodyne OPLL to stabilize the frequency difference of 1064 nm lasers to the reference frequency standard.

\begin{figure}[h]
	\centering
	\includegraphics[width=8cm]{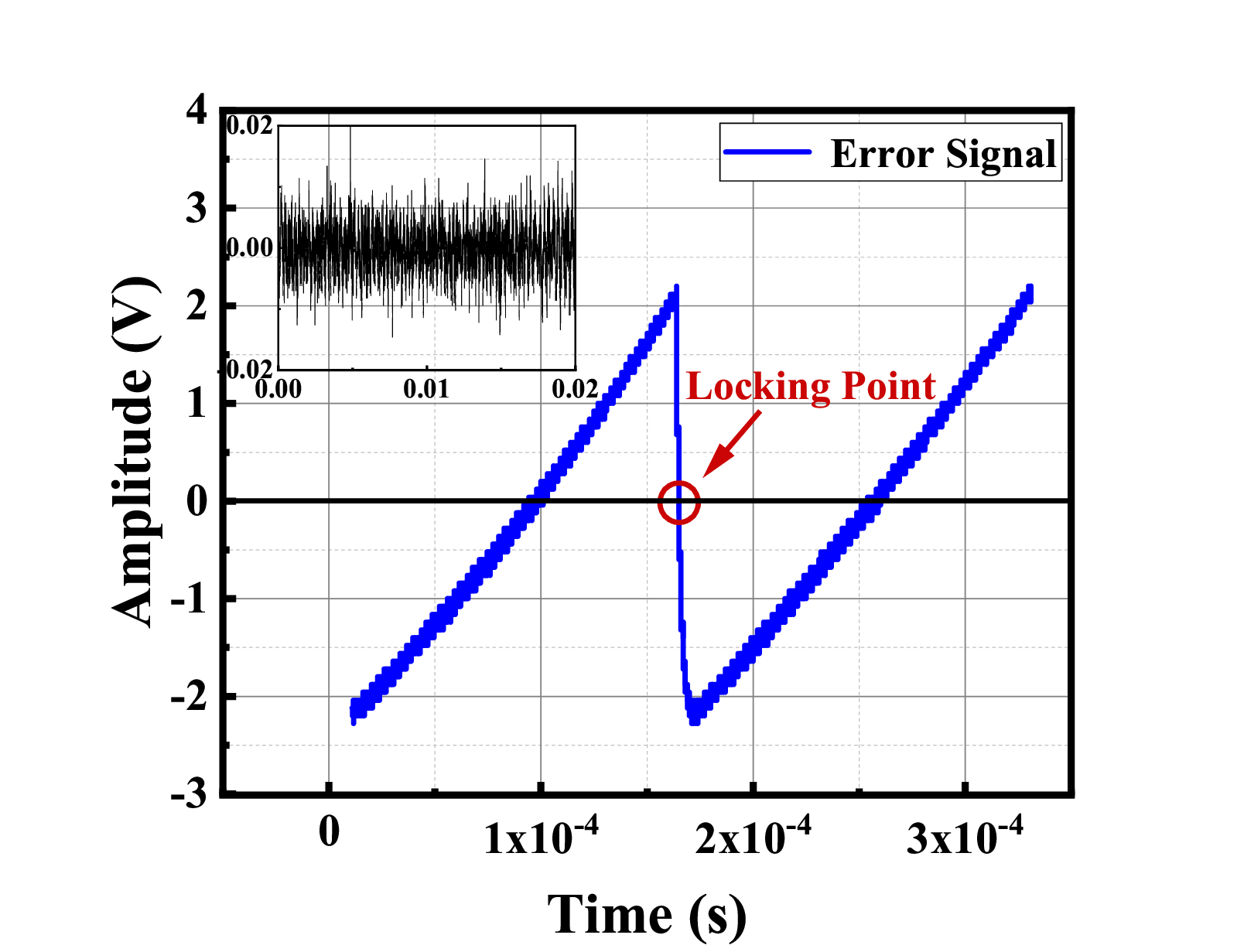}\\
	\caption{Error signal of the heterodyne optical phase locked loop between two DW-AOCs (without ramp). The red point is the lock point. Inset: Error signal after phase locking.}\label{error_signal}
\end{figure}

In the IV part, we use the OPLL technique~\cite{Optics_Communications_Volume104_Issues_4_6pp339-344_1994,J_Lightwave_Technol_17_328_342_(1999)} to synchronize the cavity-lengths change of two DW systems. In order to maintain a stable frequency offset, firstly, the beating signal is detected by the fiber coupled photodetector (PD, Thorlabs PDA8GS), and then pass through an amplifier (RF BAY, Inc. LNA-8G). Next, the amplified heterodyne beat note is imported into the offset phase lock servo (Vescent D2-135), which includes a divider, a phase-frequency detector (PFD) and a loop filter. Then, the beat note is divided by N=64 and sent to the PFD for comparing the phase and frequency of the divided-by-N beat signal with an external reference frequency. The fractional frequency stability of the reference follows that of the Cs microwave atomic clock, which is ${\rm{3}} \times {\rm{1}}{{\rm{0}}^{{\rm{ - 12}}}}$ at 1 s. By changing either the frequency of external reference or the PZT of one main-cavity, the frequency difference of 1064 nm good-cavity lasers can be precisely controlled. The PFD acts as a frequency comparator and outputs a phase lock error signal. The error signal is depicted in Fig. \ref{error_signal}, which is proportional to the phase difference between the beat note and the external reference. The inset of Fig. \ref{error_signal} is the error signal after phase locking. Finally, the error signal is fed back to the PZT of one main-cavity through proportional-integral-differential (PID) loop filter, and the synchronous change of the cavity lengths between two DW-AOCs is realized.

Note that the main-cavity of DW-AOC can also be locked to an external 1064 nm optical frequency standard~\cite{Appl_Phys_B_Laser_Opt_vol_98_pp61_67_2010,Journal_of_the_Optical_Society_of_AmericaB_30_1546-1550_(2013)}, which can further optimize the performance of the 1470 nm clock transition laser.

\section{Results and discussions}
\label{Results_and_discussions}

\subsection{Experimental results}
\begin{figure}[t]
	\subfigure[]{\includegraphics[width=0.5\textwidth]{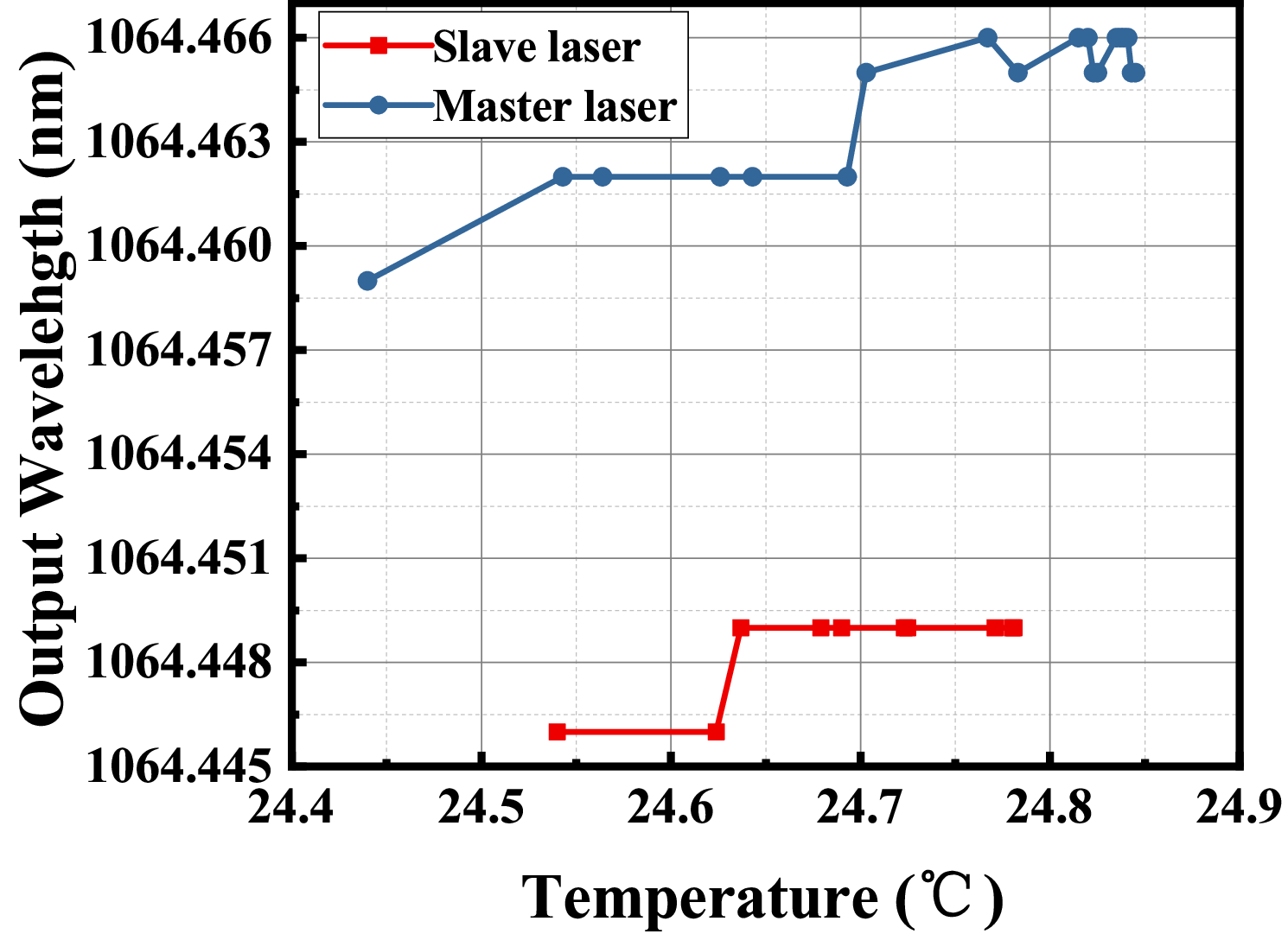}
		\label{wavelength_T} }
	\subfigure[]{\includegraphics[width=0.458\textwidth]{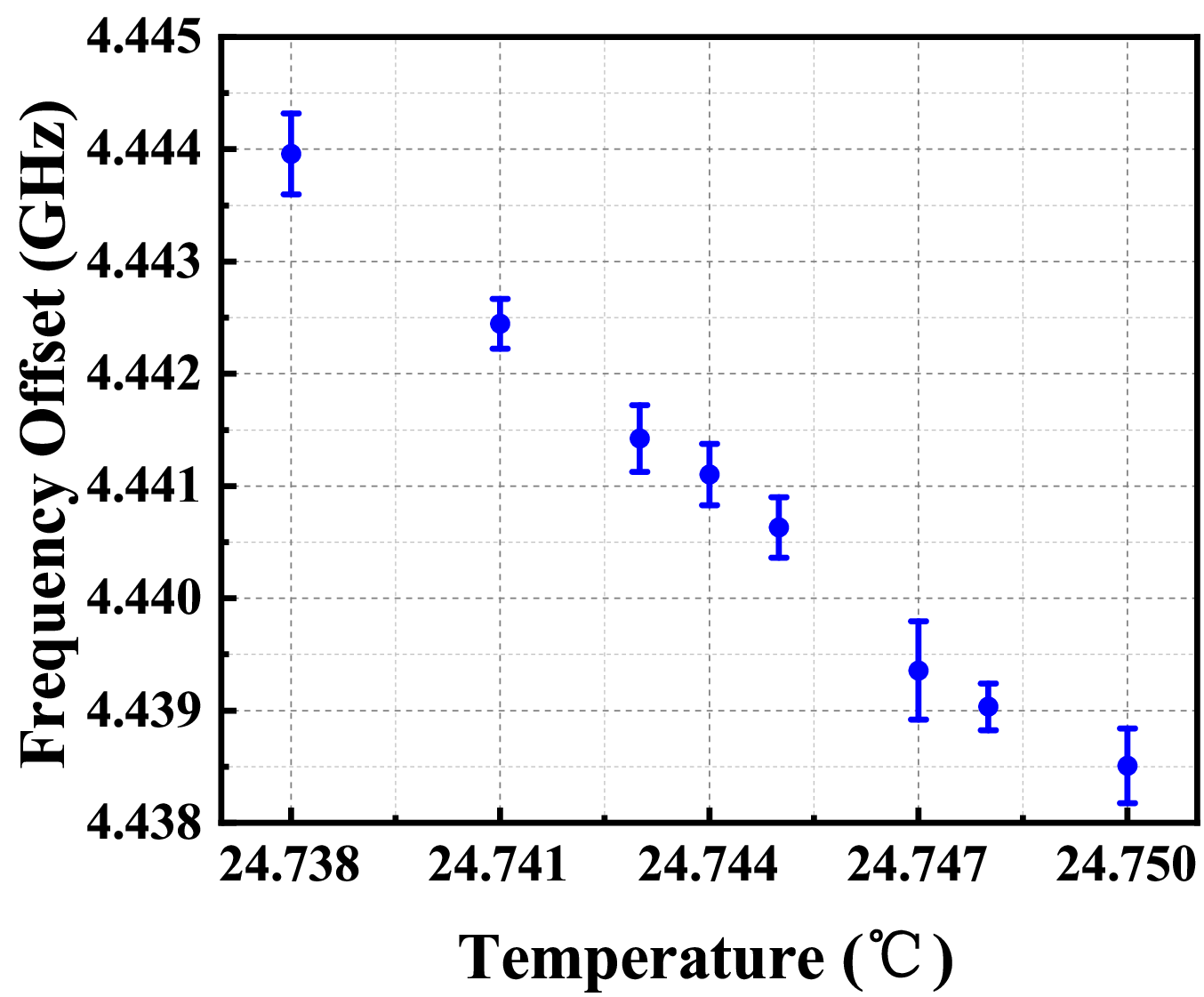}
		\label{frequency_detuning_T} }
	\caption{(a) The wavelengths of the master laser (ML) and the slave laser (SL) versus the temperatures of each Nd:YAG crystal, and the blue circular-dot line and the red squared-dot line represent the temperature characteristics of the ML and the SL, respectively. (b) The offset frequency of the two free-running 1064 nm lasers versus the crystal temperature of the ML. The crystal temperature of the SL is controlled at 24.78 $^{o}$C. The 808 nm pump powers of the ML and the SL are set at 1.52 W and 1.60 W, separately.}
\end{figure}

For simplicity, the two 1064 nm good-cavity lasers is called the master laser (ML) and the slave laser (SL), respectively. The offset frequency between the two free-running 1064 nm lasers is tunable in the range of 1-9 GHz by adjusting the temperature of each Nd:YAG crystal. The 1064 nm lasers work in the single-mode operation by setting the 808 nm pump powers, cavity lengths and crystal temperatures at appropriate values. The wavelengths of the ML and the SL versus their respective crystal temperature are shown in Fig. \ref{wavelength_T}. The two 1064 nm good-cavity lasers are both operated in mode-hop-free temperature range. The 808 nm pump power of the ML and the SL are separately set at 1.52 W and 1.60 W. The crystal temperature of the SL is controlled at 24.78 $^{o}$C, and the temperature control precision of the Nd:YAG crystal is better than 0.01 $^{o}$C. Figure \ref{frequency_detuning_T} depicts that the frequency difference of the two 1064 nm lasers changes with the crystal temperature of the ML. To obtain a stable single-mode characteristic of the 1064 nm lasers, the crystal temperatures of the ML and the SL are finally stabilized at 24.74 $^{o}$C and 24.78 $^{o}$C. As for phase locking, the servo out signal of the phase locking loop is led to the PZT driver of the SL to sweep its cavity length, aiming at obtaining an error signal with a steep slope. Setting the frequency divider to N=64 (to increase the lock time), external reference (sinusoidal signal; amplitude 12 dBm), without ramp, the frequency difference of the 1064 nm lasers is close to the reference frequency by adjusting the PZT driver of the SL or the external reference frequency.

\begin{figure}[h]
	\centering
	\includegraphics[width=8.5cm]{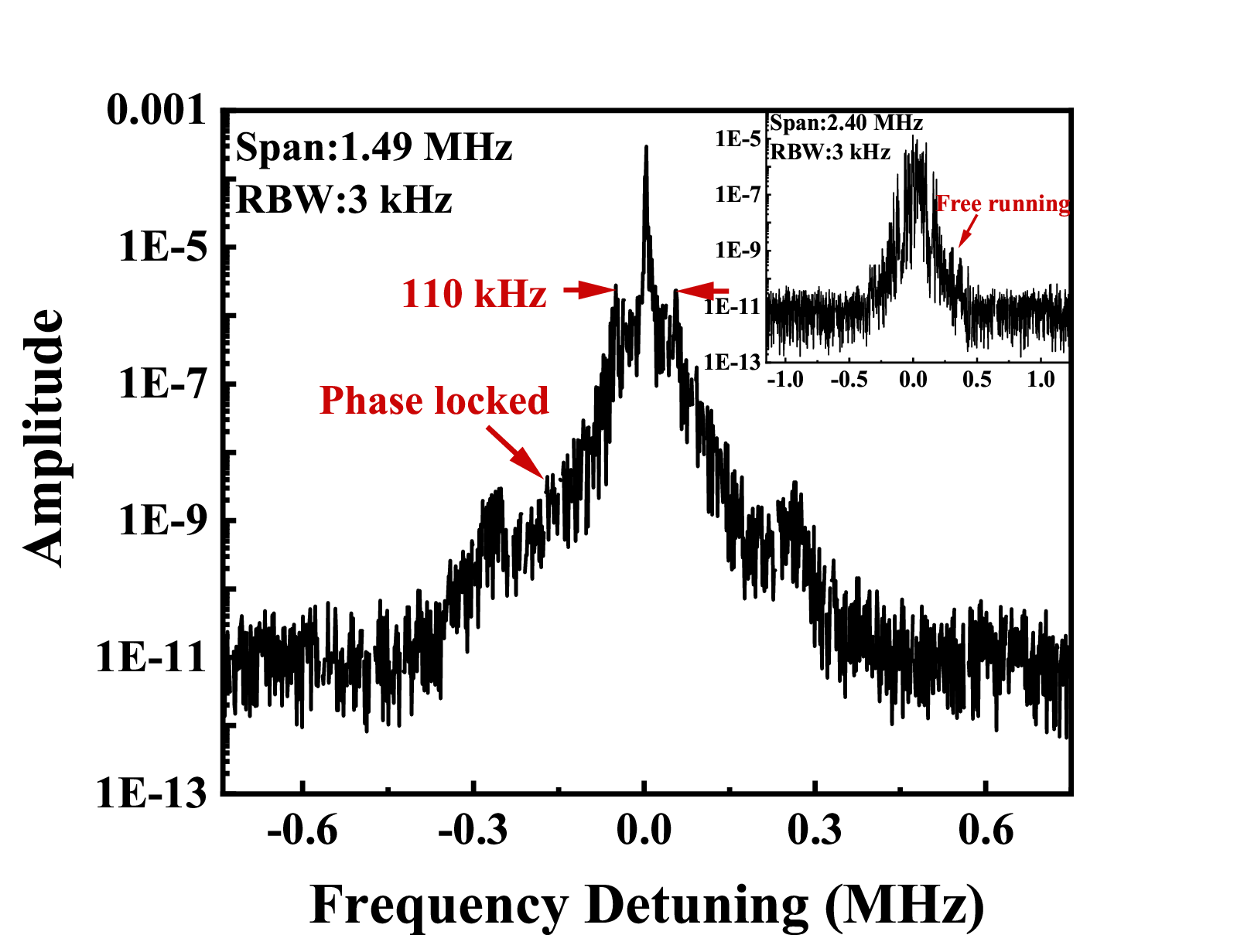}\\
	\caption{Comparison of the beat-note spectra of 1064 nm lasers with and without phase locking (RBW: 3 kHz). Inset: The beat-note spectrum of the free running 1064 nm lasers.}\label{after-before-phase-locking}
\end{figure}

\begin{figure}[h]
	\subfigure[]{\includegraphics[width=0.5\textwidth]{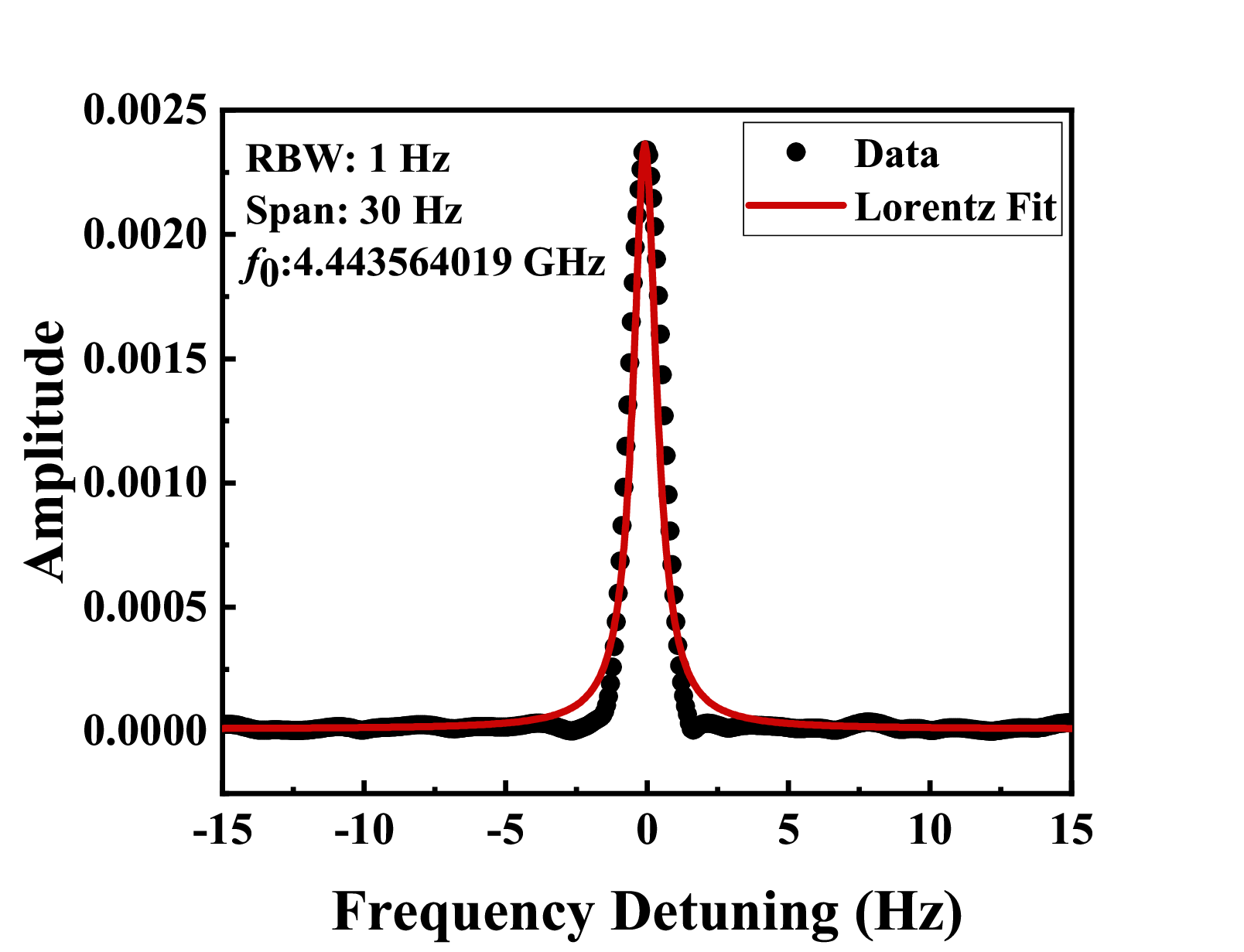}
		\label{1064_linewidth} }
	\subfigure[]{\includegraphics[width=0.5\textwidth]{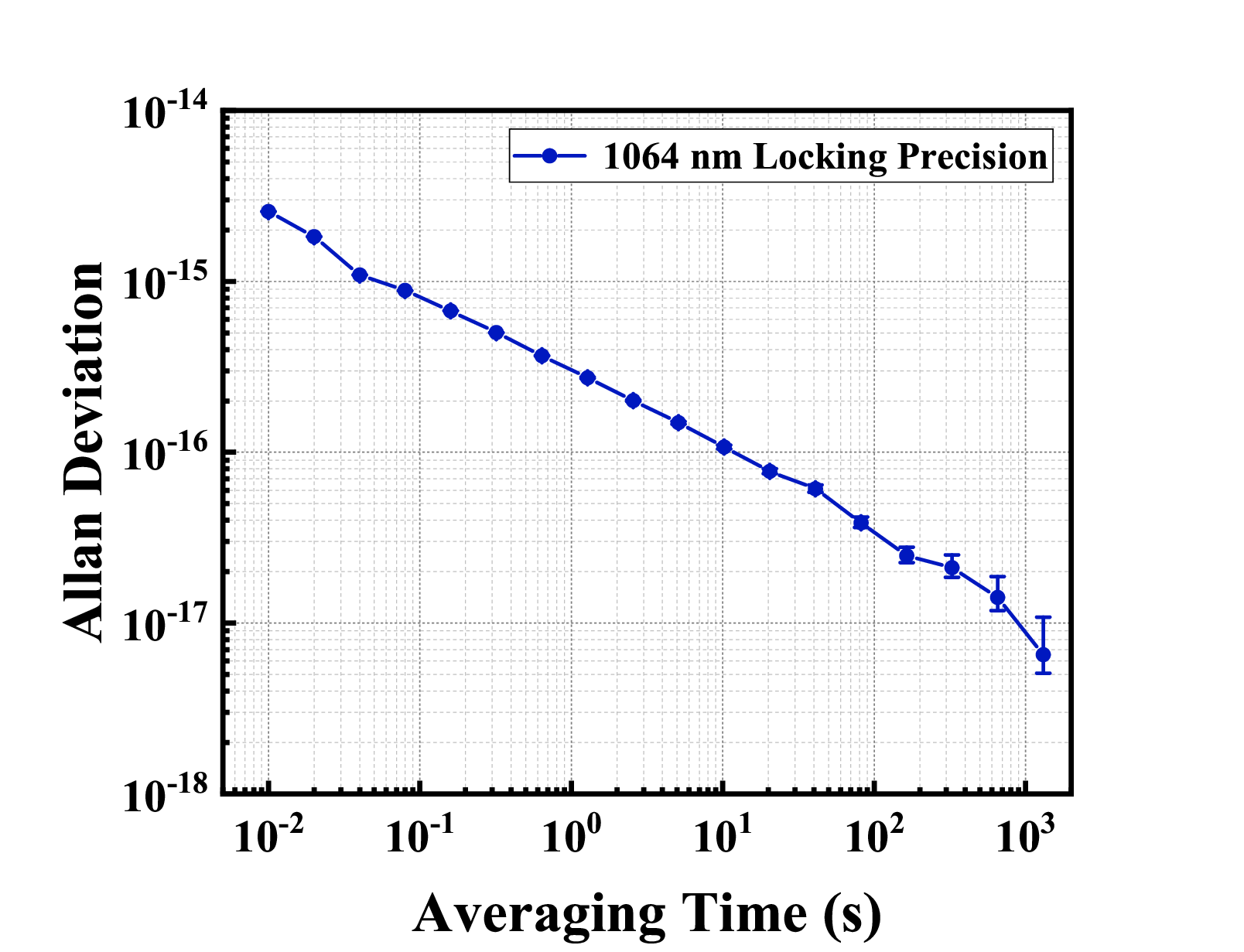}
		\label{1064_stability} }
	\caption{(a) A typical beat-note spectrum of 1064 nm good-cavity lasers after phase locking. The black points represent the original data, and the red line is the Lorentz fit of the beat note (span: 30 Hz; resolution bandwidth: 1 Hz). (b) Allan deviation of the frequency difference between two 1064 nm good-cavity lasers, it depicts that the tracking accuracy of two DW systems is ${\rm{3}} \times {\rm{1}}{{\rm{0}}^{{\rm{ - 16}}}}$ at 1 s, and can reach $1 \times {10^{ - 17}}$ at 1000 s.}
\end{figure}

We adjust the gain and the PID parameters of the phase locking loop to improve the performance of the feedback. The first step is to optimize the gain. Tuning the gain to minimize the RMS noise on the error signal in oscilloscope, and we can further fine-tune the gain by minimizing the beating linewidth in the spectrum analyzer. Figure \ref{after-before-phase-locking} depicts the spectra of the beat note acquired before and after phase locking, and the inset of Fig. \ref{after-before-phase-locking} represents the free running case. The frequency of the beat-note signal drifts about 135 kHz in 1 s without phase locking, and the whole band drifts over 6 MHz on a longer time scale. The frequency drift is stopped after phase locking and replaced by a narrow beating linewidth. The next step is to optimize the PID parameters. The PZT bandwidth of each cavity is 50 kHz without load, and the maximum dilation of each PZT is 3 $\rm{\mu}$m. The fractional frequency stability of the 1064 nm beat signal is measured in different offset lock. According to the experimental results, we finally set $\omega_{\rm{I}}$ at 4 kHz, $\omega_{\rm{D}}$ at 125 kHz and $\omega_{\rm{HF}}$ in the off position. The $\omega_{\rm{I}}$, $\omega_{\rm{D}}$ and $\omega_{\rm{HF}}$ represent the frequency where the gain switches from having integral gain to having proportional gain, the gain switches from proportional to differential and the gain begins to fall off at high frequency, respectively. After improving the feedback of the OPLL, the servo output voltage is sent to the PZT of the SL. Figure \ref{1064_linewidth} shows a beat-note spectrum of 1064 nm lasers after phase locking. The span and the resolution bandwidth (RBW) are 30 Hz and 1 Hz separately. The Lorentz fitting linewidth of the beat-note signal is 1 Hz, which is limited by the 1 Hz RBW of the spectrum analyzer (Agilent N9020A). To increases the resolution, the frequency stability of the beat-note signal is simultaneously recorded by a frequency counter. As shown in Fig. \ref{1064_stability}, the Allan deviation is calculated as a function of the gate time $\tau$ of the counter ($\tau {\rm{ = 0}}{\rm{.01 s}}$). It depicts that the frequency tracking accuracy between the two main-cavities of DW-AOCs is ${\rm{3}} \times {\rm{1}}{{\rm{0}}^{{\rm{ - 16}}}}$ at 1 s, and can reach $1 \times {10^{ - 17}}$ at 1000 s. We note that, limited by the temperature-controlling technology, the mode hopping of the 1064 nm laser occurs after the phase-locking time was up to 3 hours. Therefore the Allan deviation reaches a floor at 1000 s. The long-term stability of the 1064 nm beating signal can be optimized by improving the precision of the temperature control of the main-cavity of the DW-AOCs.

\begin{figure}[h]
	\subfigure[]{\includegraphics[width=0.5\textwidth]{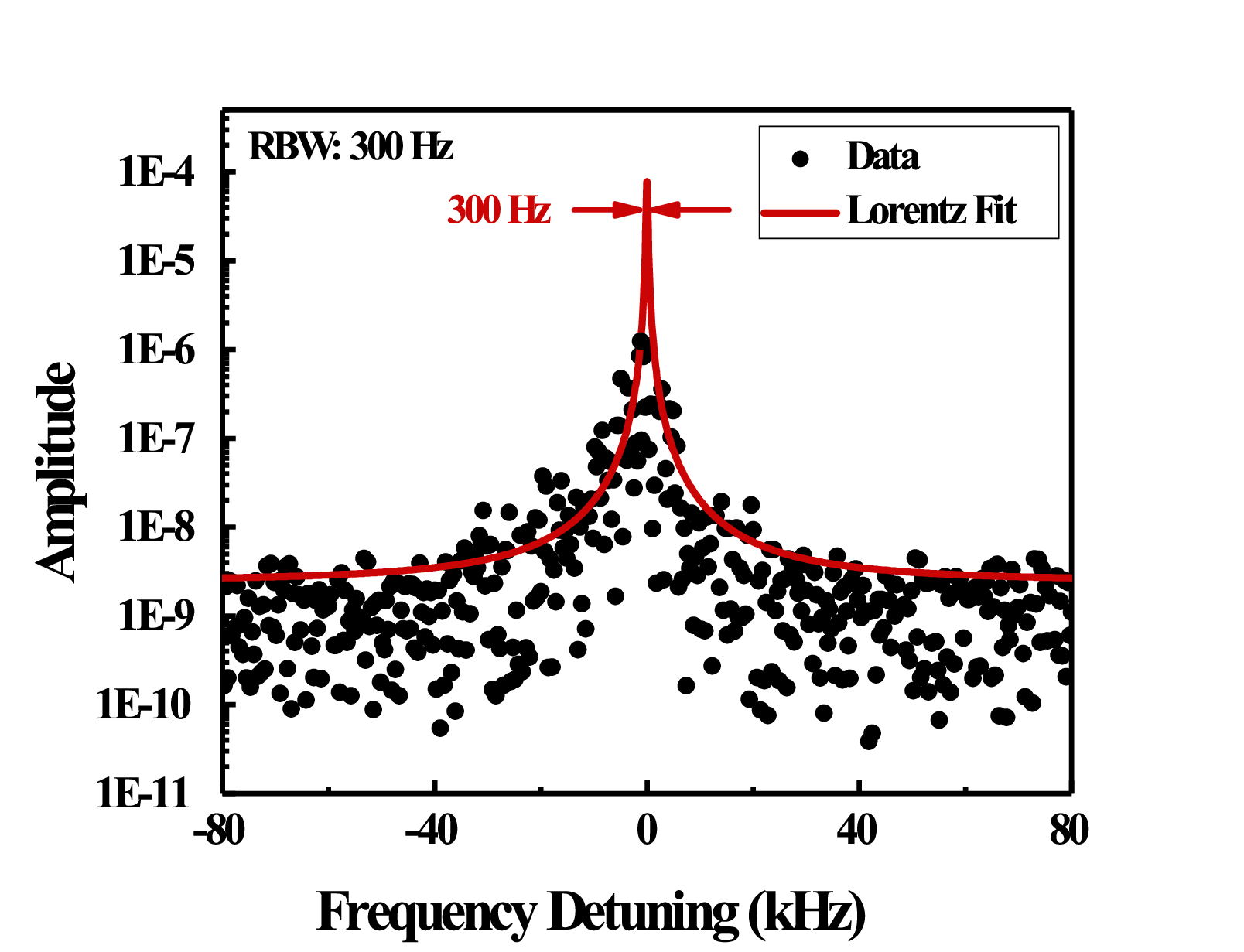}
		\label{1470_linewidth_300Hz} }
	\subfigure[]{\includegraphics[width=0.5\textwidth]{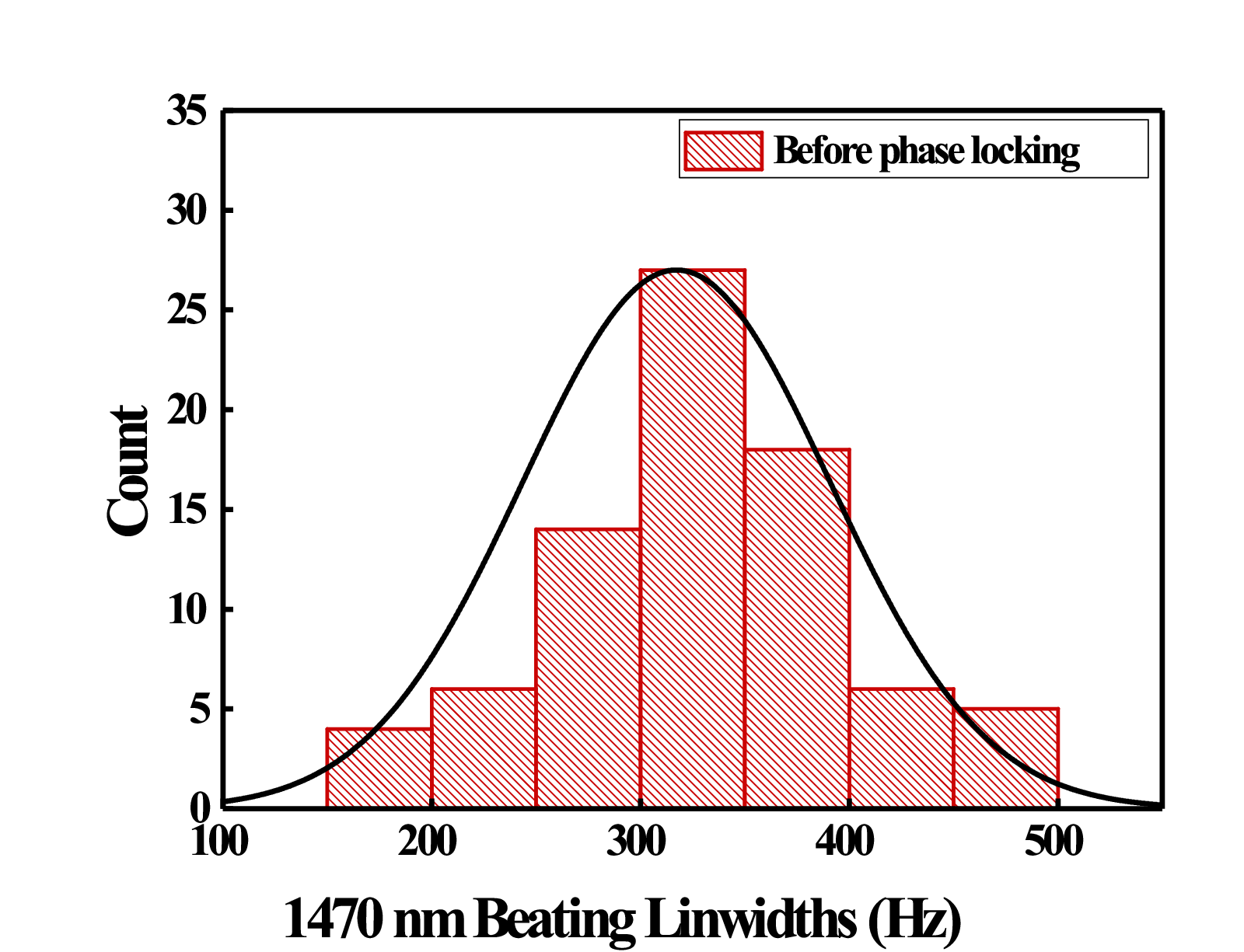}
		\label{1470nm-linewidtrh-distribution-before-phase-locking} }
	\subfigure[]{\includegraphics[width=0.5\textwidth]{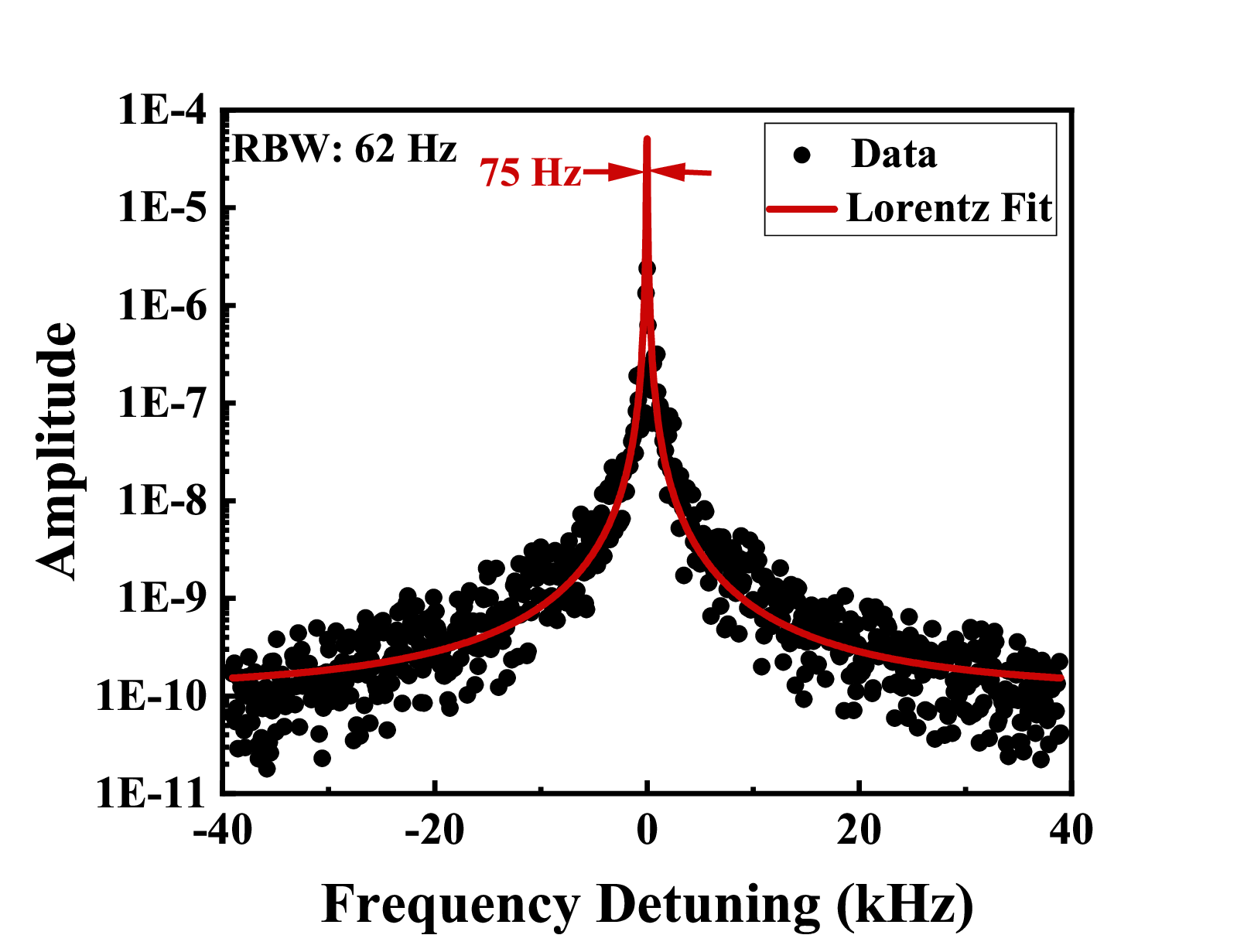}
		\label{1470_linewidth_75Hz} }
	\subfigure[]{\includegraphics[width=0.5\textwidth]{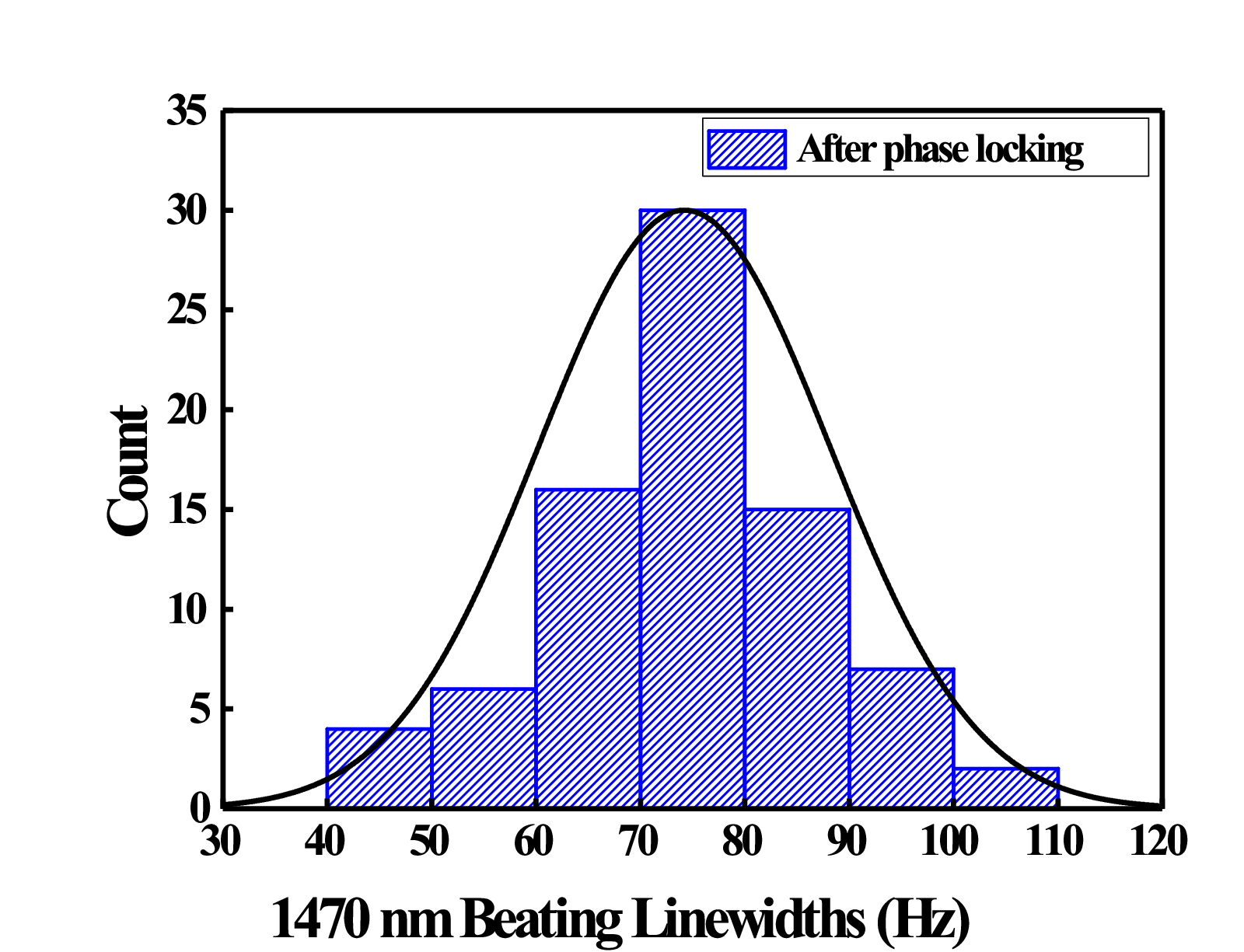}
		\label{1470nm-linewidtrh-distribution-after-phase-locking} }
	\caption{Linewidth measurements: (a) a typical beat-note spectrum of 1470 nm bad-cavity lasers before phase locking (RBW: 300 Hz), (b) linewidth distribution of 80 groups of the spectra with 50 Hz step and the Lorentz fit before phase locking, (c) a typical beat-note spectrum of 1470 nm bad-cavity lasers after phase locking (RBW: 62 Hz), (d) linewidth distribution of 80 groups of the spectra with 10 Hz step and the Lorentz fit after phase locking.}
\end{figure}

Fig.~\ref{1470_linewidth_300Hz} and \ref{1470_linewidth_75Hz} are the typical beat-note spectra of 1470 nm bad-cavity lasers before and after phase locking. The RBW of Fig.~\ref{1470_linewidth_300Hz} and \ref{1470_linewidth_75Hz} are set to be 300 Hz and 62 Hz. The RBW in these two cases are different, since the beat-note signal is still affected by the residual cavity-pulling effect before cavity-length stabilization. If we rescale the RBW, making it smaller, to measure the 1470 nm beat-note signal before phase locking, the span of the measurement will also becoming smaller. This will lead to the fact that it is difficult to flatten the two sides of the fitting line, and also result in missing data points at distant locations on both sides of the beating signal. It is hard to fit the data well with the Lorentz line shape. For this reason, we chose an appropriate RBW value, namely, RBW = 300, to measure the linewidth of the 1470 nm beating signal before phase locking. The Lorentz fitting linewidths of the power spectrum of the 1470 nm beat-note signal in Fig.~\ref{1470_linewidth_300Hz} and \ref{1470_linewidth_75Hz} are respectively 300 $\pm$ 30.1 Hz and 75 $\pm$ 8.7 Hz.

It should note that in order to increase the coincidence between the fitting and the beating data, we fit the bottom and waist data of the Lorentzian coincide with that of the power spectrum of the 1470 nm beat-note signal as much as possible. These fittings can be treated as a Lorentzian whose center frequency is distributed randomly. The data, whose absolute value of frequency detuning are greater than 4 kHz (2 kHz) in Fig.~\ref{1470_linewidth_300Hz} (\ref{1470_linewidth_75Hz}), are fitted with a Lorentzian to determine the instantaneous linewidth, which indicates the white noise level. The data, whose absolute value of frequency detuning are less than 4 kHz (2 kHz) in Fig.~\ref{1470_linewidth_300Hz} (\ref{1470_linewidth_75Hz}), are influenced by the technical frequency noise. This method is also described and applied in~\cite{Nature.48478.2012}.

In Fig. \ref{1470nm-linewidtrh-distribution-before-phase-locking} and \ref{1470nm-linewidtrh-distribution-after-phase-locking}, we obtain the linewidth distributions of the 1470 nm beat-note spectra before and after phase locking, separately, and each case contains 80 sets of data measured at different time. The most probable beating linewidth of the 1470 nm lasers are, respectively, 316 Hz and 75 Hz, which indicates that the linewidth of each 1470 nm clock transition laser is narrowed from 223 Hz to 53 Hz by assuming that each laser contributes equally to the linewidth measurements. It shows that the cavity-length stabilization has a four-fold narrowing effect on the beating linewidth of the 1470 nm lasers.

It should note that, in our paper, only the linewidth is given to quantify the instantaneous stability of the 1470 nm lasers. However, the better way of quantifying the relative stability is the Allan deviation. Since there still exists the technical noises, especially the temperature drift of the Cs atoms in long timescales, effecting the long-term stability of the 1470 nm laser, we cannot give a good result of the Allan deviation in the current work. We leave the optimization of the long-term stability for future investigations.

\subsection{Limitation factors of DW-AOC}

The common-mode noise originates from asynchronous cavity-lengths change of the two DW-AOCs is further suppressed by utilizing the phase locking technique of 1064 nm lasers. The beating linewidth of the 1470 nm bad-cavity lasers is expected to be much smaller than that of the 1064 nm good-cavity lasers after phase locking, because of the suppressed cavity-pulling effect in bad-cavity regime. However, the beating linewidth of the 1470 nm lasers is not minimized obviously, because it is still limited by other factors.

\textbf{(1) Power stability of the 459 nm pumping laser} 

\begin{figure}[h]
	\centering
	\includegraphics[width=9cm]{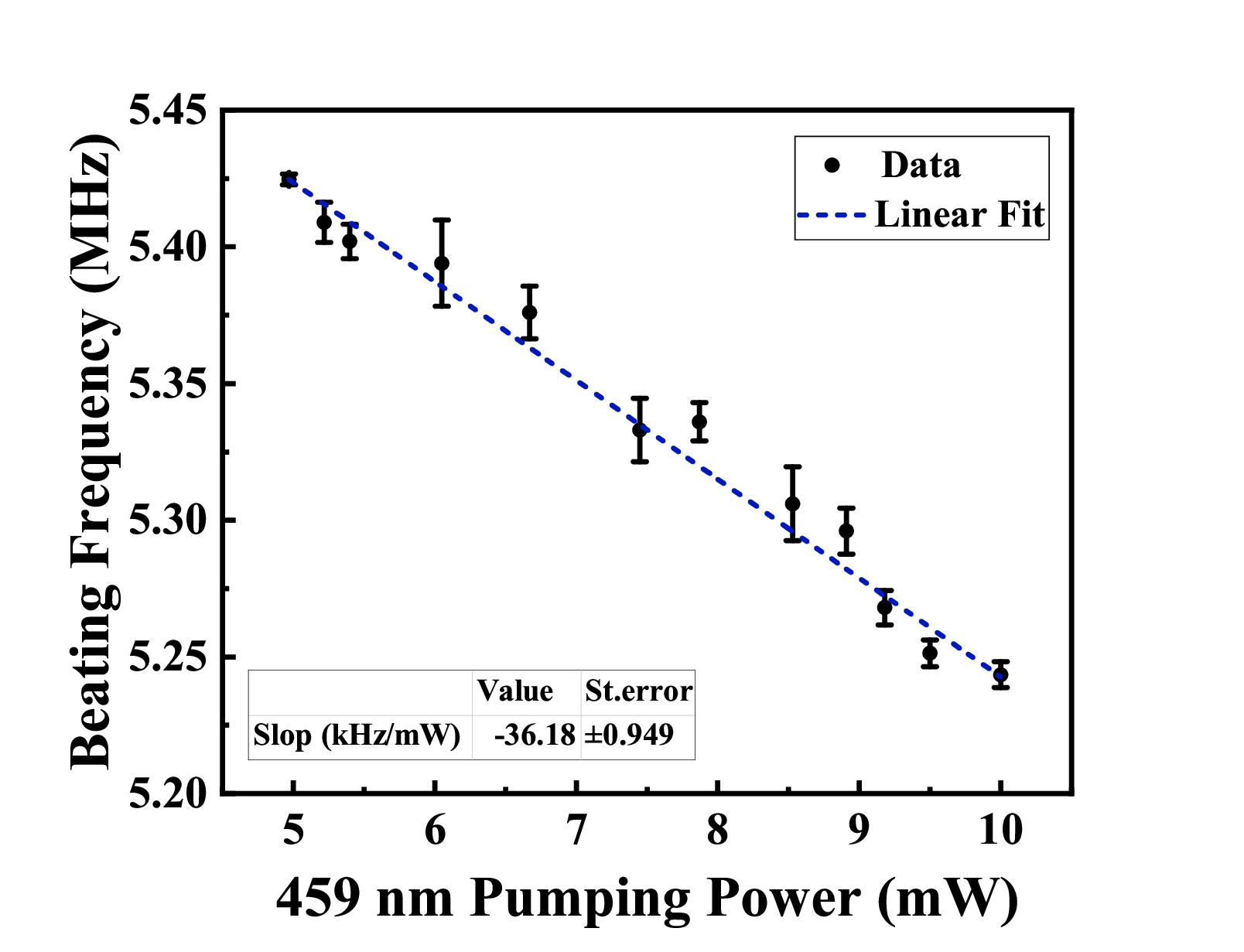}\\
	\caption{Center frequency of the 1470 nm beat-note signal depending on the 459 nm pumping power.}
	\label{Frequency_shift_of_the_1470nm_laser}
\end{figure}

In experiment, we noticed that the linewidth of the 1470 nm laser is indeed influenced by power stability of the pumping laser. To further evaluate the influence of the power stability of the pumping laser on the linewidth of the 1470 nm laser, we measured the central-frequency of the 1470 nm beat-note signal varying with the power of the 459 nm laser power, and the results are shown in Fig.~\ref{Frequency_shift_of_the_1470nm_laser}.

It can be seen that an approximately linear relationship can be obtained, with a slope of -36.1 $\pm$ 0.94 kHz/mW. The Allan deviation of the 459 nm laser power is ${\rm{1}}{\rm{.28}} \times {\rm{1}}{{\rm{0}}^{{\rm{ - 4}}}}$ at 1 s, and the pumping power is about 10 mW, which means that the linewidth broadening caused by the power fluctuation of the pumping laser is around 45 Hz. With further power stabilization of the pumping laser, we expect the linewidth broadening of the 1470 nm laser due to the power fluctuation of pumping laser can be reduced to below several hertz.

\textbf{(2) Temperature-fluctuation of the Cs cell}

The frequency shift of the 1470 nm clock transition laser caused by atomic temperature fluctuation is measured below 45 $\pm$ 1.2 kHz/$^\circ$C after cavity-length stabilization. Also, we measured the temperature of the Cs atomic vapor cell varies slowly in our system, which is, typically, below 0.1 mK in short timescales. It indicates that the impact of the atomic temperature fluctuation on the short-term stability of the 1470 nm clock-transition laser is negligible. However, in long timescales, the effect of the temperature-fluctuation on the long-term relative frequency stability of the 1470 nm clock transition laser is indeed an issue needed to be solved. To further reduce the influence of temperature-fluctuation on the long-term stability of the 1470 nm laser, our next work will put the main-cavity of the DW-AOC into a vacuum chamber.

\textbf{(3) The change of the external magnetic field}

To evaluate the impact of the magnetic-field change on the 1470 nm laser, we have measured the frequency drift of the 1470 nm beat-note signal by applying an external magnetic field to one of the systems. The central frequency of the beating signal changes linearly with the magnetic field, with a factor of about 600 kHz/G. Considering the magnetic shielding of the system, we estimate the linewidth broadening caused by the fluctuation of the external magnetic field is a few Hz, and it will be reduced greatly by designing a better magnetic shielding.  

In summary, for the instantaneous stability, the power fluctuation of the 459 nm pumping laser is the main factor resulting in the linewidth broadening of the 1470 nm laser after cavity-length stabilization in the current system. The influence of the atomic temperature change and the fluctuation of the residual magnetic field inside the cavity are relatively small. With better power stabilization of the 459 nm laser and the improvement of the temperature precision, we expect the laser linewidth to be narrowed at least one order of magnitude. However, we also note that above factors, especially the temperature drift of the Cs atomic vapor cell, and other unexpected technique noise will contribute to the long-term instability of the system. Therefore, the optimization of the above technologies is also beneficial to the long-term stability of the system.

\section{Conclusion}
\label{conclusion}

In this study, we build two DW-AOCs to analyze the linewidth characteristic of the 1064/1470 nm DW signals by optical heterodyning. Through phase locking between two independent 1064 nm good-cavity lasers, the common-mode noise caused by the asynchronous cavity-lengths change between two DW systems is suppressed. Therefore, the beating linewidth of 1470 nm bad-cavity lasers will also be narrowed after phase locking, considering that the DW lasers share a single cavity. The tracking accuracy of the two main-cavities of DW-AOCs is ${\rm{3}} \times {\rm{1}}{{\rm{0}}^{{\rm{ - 16}}}}$ at 1 s, and can reach $1 \times {10^{ - 17}}$ at 1000 s. The most probable linewidth of each 1470 nm bad-cavity laser is narrowed to 53 Hz after cavity-length stabilization. Based on this system, the influence of the limiting factors from the cavity-length noise on the performance of the AOC has been well solved. Therefore, it provides an important research platform to investigate the effects coming from the atomic ensemble, such as the the light shift, collision shift and hyperfine energy splitting, on the Cs four-level AOC. Currently, the frequency stability of the 1470 nm clock transition laser is still limited by the power stability of the 459 nm pump lasers, the temperatures fluctuation of the Cs cells and the fluctuation of the residual magnetic field. If the above problems are solved, the linewith of the 1470 nm laser is expected to be at least one order of magnitude narrower than that of the present results, which would be of great relevance for quantum metrology and new applications in fundamental and applied science, such as the study of fundamental constant variations and relativistic geodesy.

\section*{Funding}
National Natural Science Foundation of China (NSFC) (91436210).

\section*{Acknowledgments}
We acknowledge fruitful discussions with Longsheng Ma and Yanyi Jiang on the experimental part, and Yanmei Yu on the calculation of the magic wavelength.

\clearpage

%%%%%%%%%%%%%%%%%%%%%%% References %%%%%%%%%%%%%%%%%%%%%%%%%

%%%%%%%%%% If using BibTeX:
%\bibliography{sample}

\begin{thebibliography}{99}
	
	
\bibitem{Nature_506_12941_(2014)} 
B. J. Bloom, T. L. Nicholson, J. R. Williams, S. L. Campbell, M. Bishof, X. Zhang, W. Zhang, S. L. Bromley, and J. Ye, ``An optical lattice clock with accuracy and stability at the ${10^{ - 18}}$ level,'' Nature \textbf{506}, 71–75 (2014).

\bibitem{Nature_Communications_vol_6_6896_(2015)}
T. L. Nicholson, S. L. Campbell, R. B. Hutson, G. E. Marti, B. J. Bloom, R. L. McNally, W. Zhang, M. D. Barrett, M. S. Safronova, G. F. Strouse, W. L. Tew, and J. Ye, ``Systematic evaluation of an atomic clock at $2 \times {10^{ - 18}}$ total uncertainty,'' Nature Commun. \textbf{6}, 6896 (2015).

\bibitem{Nature_Photonics9_185_189(2015)}
I. Ushijima, M. Takamoto, M. Das, T. Ohkubo, and H. Katori, ``Cryogenic optical lattice clocks,'' Nature Photonics \textbf{9}, 185-189 (2015).


\bibitem{Science_358_90_94_(2017)}S. L. Campbell, R. B. Hutson, G. E. Marti, A. Goban,1 N. Darkwah Oppong, R. L. McNally, L. Sonderhouse, J. M. Robinson, W. Zhang, B. J. Bloom, and J. Ye, ``A Fermi-degenerate three-dimensional optical lattice clock,'' Science \textbf{358}, 90-94 (2017). 

\bibitem{Nature_564_87_90(2018)} W. F. McGrew, X. Zhang, R. J. Fasano, S. A. Schäffer, K. Beloy, D. Nicolodi, R. C. Brown, N. Hinkley, G. Milani, M. Schioppo, T. H. Yoon, A. D. Ludlow, ``Atomic clock performance enabling geodesy below the centimetre level,'' Nature \textbf{564}(7734), 87-90 (2018).		


\bibitem{Phys.Rev.Lett.104.070802.2010}C. W. Chou, D. B. Hume, J. C. J. Koelemeij, D. J. Wineland, and T. Rosenband, ``Frequency Comparison of Two High-Accuracy Al$^{+}$ Optical Clocks,''  Phys. Rev. Lett. \textbf{104}, 070802 (2010). 



\bibitem{Phys.Rev.Lett.116.063001.2016}N. Huntemann, C. Sanner, B. Lipphardt, Chr. Tamm, and E. Peik, ``Single-Ion Atomic Clock with $3 \times {10^{ - 18}}$ Systematic Uncertainty,'' Phys. Rev. Lett. \textbf{116}, 063001 (2016). 

\bibitem{Phys.Rev.A.99.011401(R).2019} Y. Huang, H. Guan, M. Zeng, L. Tang, and K. Gao, `` $^{40}$Ca$^+$ ion optical clock with micromotion-induced shifts below $1 \times {10^{ - 18}}$,'' Phys. Rev. A \textbf{99}, 011401(R) (2019).  



\bibitem{Metrologia_42_64_2005} R. Wynands, S. Weyers, ``Atomic fountain clocks,'' Metrologia \textbf{42}, S64-S79 (2005). 


\bibitem{Metrologia_48_283_289_2011} R. Li, K. Gibble, and K. Szymaniec, ``Improved accuracy of the NPL-CsF2 primary frequency standard: evaluation of distributed cavity phase and microwave lensing frequency shifts,'' Metrologia \textbf{48}, 283-289 (2011).

\bibitem{Metrologia_53_1123_1130_2016} J. Lodewyck, S. Bilicki, E. Bookjans, J. Robyr, C. Shi, G. Vallet, R. L. Targat, D. Nicolodi, Y. L. Coq, J. Gu\'ena, M. Abgrall, P. Rosenbusch, and S. Bize, ``Optical to microwave clock frequency ratios with a nearly continuous strontium optical lattice clock,'' Metrologia \textbf{53}, 1123-1130 (2016).


\bibitem{Metrologia_55_789_805_2018} S. Weyers, V. Gerginov, M. Kazda1, J. Rahm, B. Lipphardt, G. Dobrev, and K. Gibble, ``Advances in the accuracy, stability, and reliability of the PTB primary fountain clocks,'' Metrologia \textbf{55}, 789-805 (2018).



\bibitem{Reviews_of_Modern_Physics_2015_87(2)_637} A. D. Ludlow, M. M. Boyd, and J. Ye, ``Optical atomic clocks,'' Reviews of Modern Physics \textbf{87}, 637 (2015).


\bibitem{NaturePhotonics6_687_692(2012)} T. Kessler, C. Hagemann, C. Grebing, T. Legero, U. Sterr, F. Riehle, M. J. Martin, L. Chen, and J. Ye, ``A sub-40-mHz-linewidth laser based on a silicon single-crystal optical cavity,'' Nature Photonics \textbf{6}, 687-692 (2012).





\bibitem{Phys_Rev_Lett_vol_118_no_26_p_263202_(2017)} D. G. Matei, T. Legero, S. H\"afner, C. Grebing, R. Weyrich, W. Zhang, L. Sonderhouse, J. M. Robinson, J. Ye, F. Riehle, and U. Sterr, ``1.5 $\mu$m lasers with sub-10 mHz linewidth,'' Phys. Rev. Lett. \textbf{118}, 263202 (2017).

\bibitem{Optica.6.240.243.2019}  J. M. Robinson, E. Oelker, W. R. Milner, W. Zhang, T. Legero, D. G. Matei, F. Riehle, U. Sterr, and J. Ye,``Crystalline optical cavity at 4 K with thermal-noiselimited instability and ultralow drift,'' Optica \textbf{6}, 240-243 (2019).  







\bibitem{2005IEEE_Int_Frequency_Control_Symp}J. Chen, X. Chen, ``Optical lattice laser,'' in Proceedings of IEEE International Frequency Control Symposium(IEEE, 2005), pp. 608-610.

\bibitem{Chin_Sci_Bull_54(3)2009} J. Chen, ``Active optical clock,'' Chin. Sci. Bull., \textbf{54}, 348-352 (2009).

\bibitem{Phys.Rev.Lett.102.163601.2009}   D. Meiser, J. Ye, D. R. Carlson, and M. J. Holland. ``Prospects for a millihertz-linewidth laser,'' Phys. Rev. Lett. \textbf{102}, 163601 (2009).  



\bibitem{Nature.48478.2012} J. G. Bohnet, Z. Chen, J. M. Weiner, D. Meiser, M. J. Holland, and J. K. Thompson, ``A steady-state superradiant laser with less than one intracavity photon,'' Nature \textbf{484}, 78-81 (2012). 








\bibitem{Chin_Sci_Bull_58_pp_2033_2038_2013} T. Zhang, Y. Wang , X. Zang, W. Zhuang, and J. Chen, ``Active optical clock based on four-level quantum system,'' Chin. Sci. Bull. \textbf{58}(17), 2033-2038 (2013).

\bibitem{Opt_Lett_39_6339_(2014)} W. Zhuang and J. Chen, ``Active faraday optical frequency standard,'' Opt. Lett. \textbf{39}(21), 6339 (2014).

\bibitem{Optics_Express22_pp13269_13279(2014)} T. Maier, S. Kraemer, L. Ostermann, and H. Ritsch, ``A superradiant clock laser on a magic wavelength optical lattice,'' Optics Express \textbf{22}(11), 13269-13279 (2014).

\bibitem{IEEE_IFCS_EFTF_2015_PP363_368} D. Pan, W. Zhuang, X., X. Zhang, M. Chen, Z. Xu, and J. Chen, ``Ten years of active optical frequency standards,'' in Proceedings of IEEE International Frequency Control Symposium/European Frequency and Time Forum (IEEE, 2015), pp.363-368.

\bibitem{Phys_Rev_A_96_023412(2017)} G. A. Kazakov , J. Bohnet and T. Schumm, ``Prospects for a bad-cavity laser using a large ion crystal,'' Phys. Rev. A \textbf{96}(2), 023412 (2017).

\bibitem{Physical_Review_A96_013847(2017)} S. A. Sch$\ddot{a}$ffer, B. T. R. Christensen, M. R. Henriksen and J. W. Thomsen, ``Dynamics of bad-cavity enhanced interaction with cold Sr atoms for laser stabilization,'' Phys. Rev. A \textbf{96}(1), 013847 (2017).

\bibitem{PRX8_021036(2018)} B. Hutson, A. Goban, G. E. Marti, J. Ye, and J. K. Thompson, ``Frequency measurements of superradiance from the strontium clock transition,'' Physical Review X \textbf{8}(2), 021036 (2018).

\bibitem{IEEE_TUFFC_65PP1958_1964(2018)} D. Pan, T. Shi and J. Chen, ``Dual-wavelength good-bad-cavity laser system for cavity-stabilized active optical clock,'' IEEE Transactions on Ultrasonics, Ferroelectrics, and Frequency Control \textbf{65}, 1958-1964 (2018).

\bibitem{IEEE_IFCS_2014pp242_245} D. Pan, Z. Xu, X. Xue, W. Zhuang, and J. Chen, ``Lasing of cesium active optical clock with 459 nm laser pumping,'' in Proceedings of IEEE International Frequency Control Symposium (IEEE, 2014), pp. 242-245.

\bibitem{Chin_Phys_Lett_32_083201(2015)} Z. Xu, D. Pan, W. Zhuang, and J. Chen, ``Experimental Scheme of 633 nm and 1359 nm good-bad cavity dual-wavelength active optical frequency standard,'' Chin. Phys. Lett. \textbf{32}(8), 083201 (2015).

\bibitem{Rev_Sci_Inst_89_043102(2018)} T. Shi, D. Pan, P. Chang, H. Shang, and J. Chen, ``A highly integrated single-mode 1064 nm laser with 8.5 kHz linewidth for dual-wavelength active optical clock,'' Rev. Sci. Inst. \textbf{89}(4), 043102 (2018).

\bibitem{Appl_Phys_B_vol_60_S241_S248_1995} M. Prevedelli, T. Freegarde, and T.W. H$\ddot{a}$nsch, ``Phase locking of grating-tuned diode lasers,'' Appl. Phys. B \textbf{60}, S241-S248 (1995).

\bibitem{Science_288_635_699_(2000)} D. J. Jones, S. A. Diddams, J. K. Ranka, A. Stentz, R. S. Windeler, J. L. Hall, and S. T. Cundiff, ``Carrier-envelope phase control of femtosecond mode-locked lasers and direct optical frequency synthesis,'' Science \textbf{288}(5466), 635-699 (2000).

\bibitem{Optics_Letters_34_2958_2960_(2009)}P. Khosropanah, A. Baryshev, W. Zhang, W. Jellema, J. N. Hovenier, J. R. Gao, T. M. Klapwijk, D. G. Paveliev, B. S. Williams, S. Kumar, Q. Hu, J. L. Reno, B. Klein, and J. L. Hesler, ``Phase locking of a 2.7 THz quantum cascade laser to a microwave reference,'' Opt. Lett. \textbf{34}(19), 2958-2960 (2009).

\bibitem{OpticsExpress_vol18_pp8621-8629_2010} F. Friederich, G. Schuricht, A. Deninger, F. Lison, G. Spickermann, P. H. Bolívar, and H. G. Roskos, ``Phase-locking of the beat signal of two distributed-feed-back diode lasers to oscillstors working in the MHz to THz range,'' Opt. Express \textbf{18}(8), 8621-8629 (2010).

\bibitem{IEEE_Photonics_Technology_Letters_Vol_24_2012} M. R. H. Khan, E. H. Bernhardi, D. A. I. Marpaung, M. Burla,
R. M. de Ridder, K. W\"orhoff, M. Pollnau, and C. G. H. Roeloffzen, ``Dual-frequency distributed feedback laser with optical frequency locked loop for stable microwave signal generation,'' IEEE Photonics Technology Lett. \textbf{24}(16), 1431-1433 (2012).

\bibitem{Appl_Phys_B_vol_123_no_9_2017} M. Lipka, M. Parniak, and W. Wasilewski, ``Optical frequency locked loop for long-term stabilization of broad-line DFB lasers frequency difference,'' Appl. Phys. B \textbf{123}(9), 238 (2017).

\bibitem{Phys.Rev.A.94.012505.2016}  M. S. Safronova, U. I. Safronova, and C. W. Clark, ``Magic wavelengths, matrix elements, polarizabilities, and lifetimes of Cs,'' Phys. Rev. A \textbf{94}(1), 012505 (2016). 	




\bibitem{Optics_Communications_Volume104_Issues_4_6pp339-344_1994}G. Santarelli, A. Clairon, S.N. Lea, and G.M. Tino, ``Heterodyne optical phase-locking of extended-cavity semiconductor lasers at 9 GHz,'' Opt. Commun. \textbf{104}, 339-344 (1994).

\bibitem{J_Lightwave_Technol_17_328_342_(1999)}A. C. Bordonalli, C.Walton, and A. J. Seeds, ``High-performance phase locking of wide linewidth lasers by combined use of optical injection locking and optical phase-lock loop,'' J. Lightwave Technol. \textbf{17}(2), 328-342 (1999).

\bibitem{Appl_Phys_B_Laser_Opt_vol_98_pp61_67_2010} Y. Jiang, S. Fang, Z. Bi, X. Xu, and L. Ma, ``Nd:YAG lasers at 1064 nm with 1-Hz linewidth,'' Appl. Phys. B, Laser Opt. \textbf{98}, 61-67 (2010).

\bibitem{Journal_of_the_Optical_Society_of_AmericaB_30_1546-1550_(2013)} H. Chen,  Y. Jiang,  S. Fang, Z. Bi, and L. Ma, ``Frequency stabilization of Nd:YAG lasers with most probable linewidth of 0.6 Hz,'' Journal of the Optical Society of America B \textbf{30}(6), 1546-1550 (2013).	

	
	
	
	
	
	
	
	
	
	
	
	
\end{thebibliography}

%%%%%%%%%% If preparing manually:
% \newcommand{\enquote}[1]{``#1''}

% \bibitem{Zhang:14}
% Y.~Zhang, S.~Qiao, L.~Sun, Q.~W. Shi, W.~Huang, L.~Li, and Z.~Yang,
%   \enquote{Photoinduced active terahertz metamaterials with nanostructured
%   vanadium dioxide film deposited by sol-gel method,}
%   {\protect\JournalTitle{Optics Express}} \textbf{22}, 11070--11078 (2014).

% {Optical Society}, \enquote{{OSA Publishing},}
%   \url{http://www.osapublishing.org}.

% \bibitem{FORSTER2007}
% P.~Forster, V.~Ramaswamy, P.~Artaxo, T.~Bernsten, R.~Betts, D.~Fahey,
%   J.~Haywood, J.~Lean, D.~Lowe, G.~Myhre, J.~Nganga, R.~Prinn, G.~Raga,
%   M.~Schulz, and R.~V. Dorland, \enquote{Changes in atmospheric consituents and
%   in radiative forcing,} in \enquote{Climate Change 2007: The Physical Science
%   Basis. Contribution of Working Group 1 to the Fourth assesment report of
%   Intergovernmental Panel on Climate Change,}  S.~Solomon, D.~Qin, M.~Manning,
%   Z.~Chen, M.~Marquis, K.~B. Averyt, M.~Tignor, and H.~L. Miler, eds.
%   (Cambridge University Press, 2007).

% \end{thebibliography}

\end{document}